\documentclass[letterpaper,twocolumn,prl,aps,superscriptaddress,floatfix]{revtex4}

\usepackage[latin9]{inputenc}
\usepackage{amsmath}
\usepackage{amssymb}
\usepackage{graphicx}
\usepackage{esint}
\usepackage{times}
\usepackage{enumitem}
\usepackage[markup=underlined]{changes}

\usepackage[unicode=true,
bookmarks=true,bookmarksnumbered=false,bookmarksopen=false,
breaklinks=false,pdfborder={0 0 1},backref=false,colorlinks=true]
{hyperref}
\hypersetup{
    linkcolor=magenta, urlcolor=blue, citecolor=blue, pdfstartview={FitH}, hyperfootnotes=false, unicode=true}

\makeatletter

\pdfpageheight\paperheight
\pdfpagewidth\paperwidth

\@ifundefined{textcolor}{}
{%
    \definecolor{BLACK}{gray}{0}
    \definecolor{WHITE}{gray}{1}
    \definecolor{RED}{rgb}{1,0,0}
    \definecolor{GREEN}{rgb}{0,1,0}
    \definecolor{BLUE}{rgb}{0,0,1}
    \definecolor{CYAN}{cmyk}{1,0,0,0}
    \definecolor{MAGENTA}{cmyk}{0,1,0,0}
    \definecolor{YELLOW}{cmyk}{0,0,1,0}
}

\usepackage{xcolor}
\usepackage{soul}

\newcommand{\bra}[1]{\ensuremath{\left\langle#1\right|}}
\newcommand{\ket}[1]{\ensuremath{\left|#1\right\rangle}}
\definecolor{blue}{rgb}{0,0,1}
\definecolor{red}{rgb}{1,0,0}
\definecolor{green}{rgb}{0,1,0}

\usepackage{soul}

\makeatother

\begin{document}
\title{A biased-erasure cavity qubit with hardware-efficient quantum error detection}

\author{Jiasheng Mai}
\thanks{These authors contributed equally to this work.}
\affiliation{Southern University of Science and Technology, Shenzhen 518055, China}
\affiliation{International Quantum Academy, Shenzhen 518048, China}

\author{Qiyu Liu}
\thanks{These authors contributed equally to this work.}
\affiliation{International Quantum Academy, Shenzhen 518048, China}

\author{Xiaowei Deng}
\thanks{These authors contributed equally to this work.}
\email{dengxiaowei@iqasz.cn}
\affiliation{International Quantum Academy, Shenzhen 518048, China}

\author{Yanyan Cai}
\affiliation{Southern University of Science and Technology, Shenzhen 518055, China}
\affiliation{International Quantum Academy, Shenzhen 518048, China}

\author{Zhongchu Ni}
\affiliation{International Quantum Academy, Shenzhen 518048, China}

\author{Libo Zhang}
\affiliation{Southern University of Science and Technology, Shenzhen 518055, China}
\affiliation{International Quantum Academy, Shenzhen 518048, China}

\author{Ling Hu}
\affiliation{International Quantum Academy, Shenzhen 518048, China}
\affiliation{Shenzhen Branch, Hefei National Laboratory, Shenzhen 518048, China}

\author{Pan Zheng}
\affiliation{International Quantum Academy, Shenzhen 518048, China}

\author{Song Liu}
\affiliation{International Quantum Academy, Shenzhen 518048, China}
\affiliation{Shenzhen Branch, Hefei National Laboratory, Shenzhen 518048, China}

\author{Yuan Xu}
\email{xuyuan@iqasz.cn}
\affiliation{International Quantum Academy, Shenzhen 518048, China}
\affiliation{Shenzhen Branch, Hefei National Laboratory, Shenzhen 518048, China}
\author{Dapeng Yu}
\affiliation{International Quantum Academy, Shenzhen 518048, China}
\affiliation{Shenzhen Branch, Hefei National Laboratory, Shenzhen 518048, China}

\begin{abstract}
\textbf{Erasure qubits are beneficial for quantum error correction due to their relaxed threshold requirements. While dual-rail erasure qubits have been demonstrated with a strong error hierarchy in circuit quantum electrodynamics, biased-erasure qubits---where erasures originate predominantly from one logical basis state---offer further advantages. Here, we realize a hardware-efficient biased-erasure qubit encoded in the vacuum and two-photon Fock states of a single microwave cavity. The qubit exhibits an erasure bias ratio of over 265. By using a transmon ancilla for logical measurements and mid-circuit erasure detections, we achieve logical state assignment errors below 1\% and convert over 99.3\% leakage errors into detected erasures. After postselection against erasures, we achieve effective logical relaxation and dephasing rates of $(6.2~\mathrm{ms})^{-1}$ and $(3.1~\mathrm{ms})^{-1}$, respectively, which exceed the erasure error rate by factors of 31 and 15, establishing a strong error hierarchy within the logical subspace. These postselected error rates indicate a coherence gain of about 6.0 beyond the break-even point set by the best physical qubit encoded in the two lowest Fock states in the cavity. Moreover, randomized benchmarking with interleaved erasure detections reveals a residual logical gate error of 0.29\%. This work establishes a compact and hardware-efficient platform for biased-erasure qubits, promising concatenations into outer-level stabilizer codes toward fault-tolerant quantum computation.} 
\end{abstract}

\maketitle
\vskip 0.5cm

\noindent \textbf{\large{}Introduction}{\large\par}
\noindent Quantum error correction (QEC) is essential for realizing practical and scalable quantum computation~\cite{Nielsen2010}, and has recently been demonstrated to extend the quantum information lifetime beyond the break-even point~\cite{ofek2016,ni2023,sivak2023,google2025,Brock2025,Sun2025,Ni2025,Shruti2025}. A promising route to advance QEC lies in engineering quantum systems with biased noise structures, where errors predominantly occur along specific channels. Such structured noise can increase error thresholds and reduce resource overheads when embedded in tailored QEC codes designed to exploit this bias~\cite{stephens2013,webster2015,tuckett2019,bonilla2021,xu2023,claes2023,roffe2023}. A prominent example is the biased Pauli noise in bosonic cat qubits~\cite{grimm2020,lescanne2020}, which exhibit exponentially suppressed bit-flip errors at the cost of linearly increased phase-flip errors. This noise bias can be leveraged through concatenation with outer stabilizer codes to improve QEC performance~\cite{guillaud2019,darmawan2021,chamberland2022,putterman2025}. 

An alternative strategy is the use of erasure qubits~\cite{grassl1997,gu2025, kubica2023}, wherein the dominant error is leakage out of the logical subspace. Such leakage events constitute erasure errors that occur in a known qubit and at a known time. When these qubits are integrated into higher-level stabilizer codes, erasure errors can be detected and corrected more efficiently than Pauli errors, thereby enabling higher QEC thresholds and lower logical error rates~\cite{wu2022,kubica2023}. To fully exploit these advantages, an erasure qubit should efficiently convert leakage into detectable erasures, maintain a high ratio of erasure errors to residual errors within the logical subspace, and introduce minimal additional errors during erasure detection. In recent years, erasure qubits have been implemented in various physical platforms, such as neutral atoms~\cite{wu2022,ma2023,scholl2023}, trapped ions~\cite{kang2023}, and superconducting circuits~\cite{teoh2023,chou2024,levine2024,koottandavida2024,degraaf2025,mehta2025,huang2025}.

\begin{figure*}[t]
    \centering
    \includegraphics{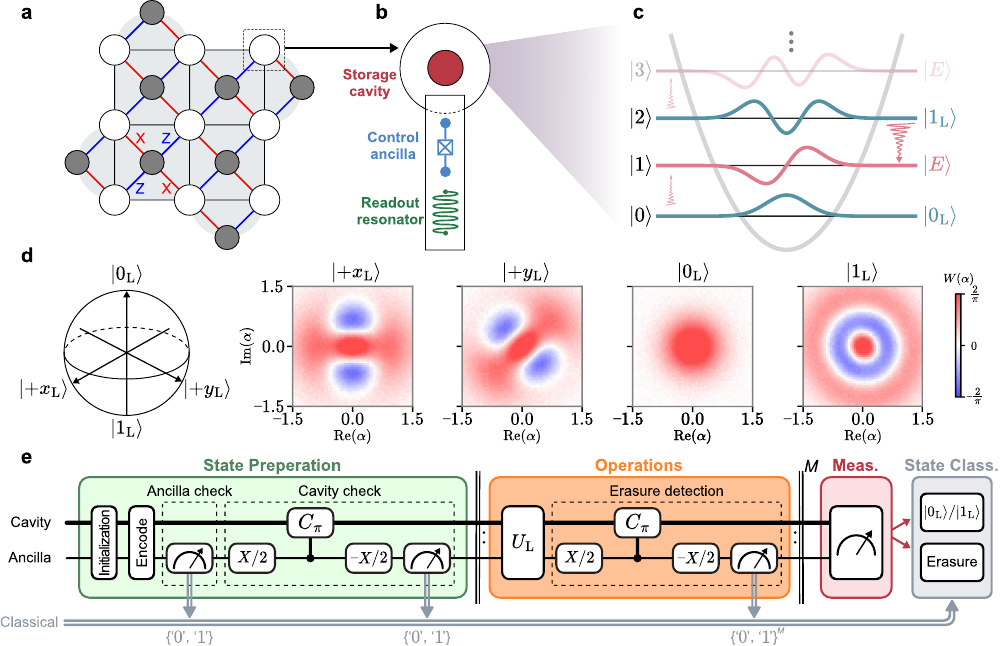}
    \caption{\textbf{Experimental scheme of the biased-erasure 02 qubit.} 
    \textbf{a.} Schematic of an XZZX surface code using biased-erasure data qubits (white circles) and stabilizer measurement qubits (grey circles). The biased noise structure of the erasure qubits can be harnessed to improve QEC performance.  
    \textbf{b.} Device schematic, consisting of a 3D coaxial stub cavity (red) for encoding the erasure qubit, a transmon ancilla (blue), and a readout resonator (green) for assisting the logical control and erasure detection.  
    \textbf{c.} Energy-level diagram of the cavity mode. The logical basis states of the erasure qubit are $\ket{0_\mathrm{L}} = \ket{0}$ and $\ket{1_\mathrm{L}} = \ket{2}$ (green). Single-photon loss and gain errors would bring the logical states into leakage states $\ket{1}$ and $\ket{3}$, which can be detected and flagged as erasures.  
    \textbf{d.} Logical Bloch sphere representation of the 02 qubit, with four cardinal states visualized using measured Wigner functions in phase space.  
    \textbf{e.} Quantum circuit for state preparation, logical operations, and measurement of the 02 erasure qubit. State preparation is performed via QOC pulses, followed by ancilla and cavity checks to purify the initialization. Logical gates are also implemented using QOC pulses, interleaved with mid-circuit erasure detection. Final end-of-line logical measurement and state classification distinguish $\ket{0_\mathrm{L}}$, $\ket{1_\mathrm{L}}$, and the erasure state $\ket{E}$ based on all prior check outcomes. 
    }
    \label{fig1}
\end{figure*}

Beyond these realizations, the concept of biased-erasure qubits---in which erasure errors originate predominantly from a single logical basis state---has been proposed to further relax threshold requirements when incorporated into the XZZX surface code~\cite{bonilla2021} (see Fig.~\ref{fig1}a) that leverages such biased erasure characteristics~\cite{sahay2023}. A hardware-efficient example of this biased-erasure qubit is the 0N qubit~\cite{sabapathy2018}, which is encoded using the vacuum and $N$th Fock states in a single harmonic oscillator by leveraging its high-dimensional Hilbert space. In circuit quantum electrodynamics (cQED) systems~\cite{Blais2021}, this 0N erasure qubit can be encoded in a single high-quality microwave cavity, where the erasure error originating from the $N$th Fock state, due to dominant photon loss, is much larger than that from the vacuum state, thereby yielding a biased-erasure qubit~\cite{sahay2023,teoh2023thesis,mori2024} with reduced hardware overhead compared to previous dual-rail encoding using multiple qubits or cavities~\cite{chou2024,levine2024,koottandavida2024}. Despite these appealing features, the experimental demonstration of the biased-erasure qubit remains elusive.

In this work, we experimentally realize such a biased-erasure qubit---the 02 qubit---encoded in a single three-dimensional (3D) superconducting microwave cavity. The demonstrated 02 erasure qubit exhibits a large erasure bias ratio: erasure error originating from one logical basis state exceeds that from the other one by about 265 times. Using end-of-line measurements and mid-circuit erasure detections with the assistance of a transmon ancilla, we achieve an average logical assignment error below 1\% and successfully convert over 99.3\% leakage errors into detected erasures. With mid-circuit erasure detections and postselection against erasures, we measure a logical relaxation rate of $(6.2~\mathrm{ms})^{-1}$ and a dephasing rate of $(3.1~\mathrm{ms})^{-1}$, substantially lower than the erasure error rate $(0.2~\mathrm{ms})^{-1}$ by factors of 31 and 15, respectively. These postselected erasure rates indicate a coherence gain factor of about 6.0 over the best physical qubit encoded in the two lowest Fock states in the cavity. Moreover, using randomized benchmarking interleaved with mid-circuit erasure detections, we characterize single-qubit logical gates with a residual gate error of $2.9\times 10^{-3}$, which is approximately 16 times lower than the erasure probability per gate. 

\begin{figure*}[t]
    \centering
    \includegraphics{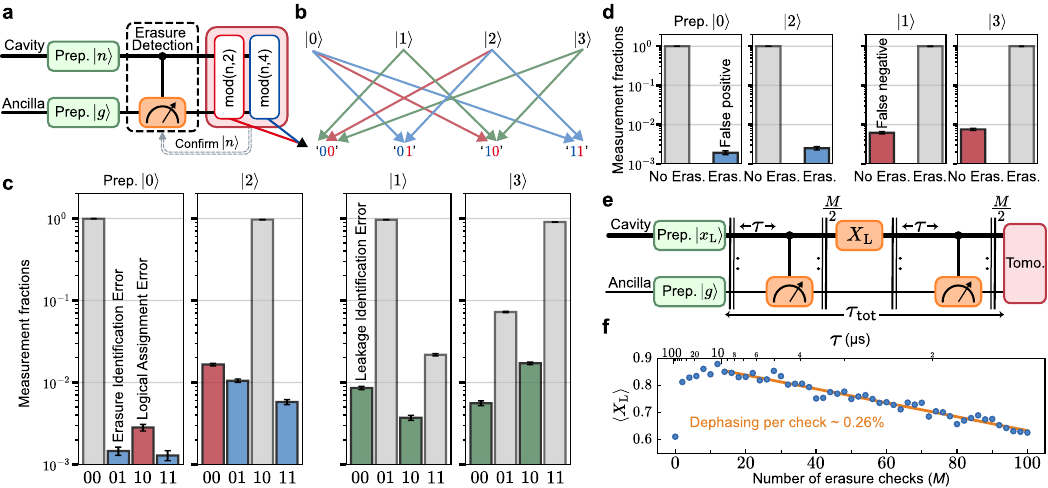}
    \caption{\textbf{Characterization of erasure detection and logical measurement.}
    \textbf{a.} Quantum circuit for quantifying logical assignment errors using cascaded photon-number modulo parity measurements (Supplementary Sec.~II-B). The dashed box indicates the interleaved mid-circuit erasure detection used to characterize false-positive and false-negative probabilities.
    \textbf{b.} Error map between the prepared logical states (top) and the measured bitstrings (bottom) of the two parity measurements in \textbf{a}. Colored arrows denote different error types: logical assignment errors (red), erasure errors (blue), and leakage detection errors (green).
    \textbf{c.} Assigned measurement outcome fractions for prepared logical states $\ket{0}$ and $\ket{2}$ (left) and leakage states $\ket{1}$ and $\ket{3}$ (right) without mid-circuit erasure detection.
    \textbf{d.} Assigned erasure fractions of the mid-circuit erasure detection conditioned on the final logical measurement outcome when initially preparing logical states~(left) and leakage states~(right), respectively. The average false-positive and false-negative fractions are $(0.22 \pm 0.02)\%$ and $(0.69 \pm 0.03)\%$, respectively. Error bars represent $\pm 1 \sigma$ standard error. 
    \textbf{e.} Experimental sequence for characterizing detection-induced dephasing. After preparing $\ket{x_\mathrm{L}}$, a variable number $M$ (even integer) of erasure checks are evenly inserted into an echoed sequence with a fixed total duration $\tau_{\mathrm{tot}} = 150$~\textmu$\mathrm{s}$; phase coherence of the logical state is extracted via Wigner tomography. 
    \textbf{f.} Extracted phase coherence $\langle X_{\mathrm{L}}\rangle$ as a function of $M$. The coherence initially improves and then decays linearly with $M$, yielding an induced dephasing error of 0.26\% per erasure detection.
    }
    \label{fig2}
\end{figure*}

\noindent \textbf{\large{}Results}{\large\par}
\noindent \textbf{Experimental scheme}{\large\par}

\noindent Figure~\ref{fig1}b illustrates our experimental device, comprising a single 3D coaxial stub cavity~\cite{reagor2016} for storing the encoded erasure qubit, a transmon~\cite{koch2007} ancilla for implementing erasure detection and logical gates, and a stripline readout resonator~\cite{axline2016} for measuring the ancilla and logical qubit states. The transmon ancilla is dispersively coupled to both the storage cavity and the readout resonator, with measured dispersive shifts of 1.69~MHz and 1.01~MHz, respectively. The storage cavity exhibits a single-photon lifetime $T_{1,c} = 0.47$~ms ($\kappa_c/2\pi=0.33$~kHz), a Ramsey dephasing time $T_{2R,c} = 0.74$~ms, and a thermal excitation population $n_{\mathrm{th},c}=0.72\%$, indicating that single-photon loss constitutes the dominant error channel in the cavity (see Supplementary Sec.~I-B for detailed parameters).

The bosonic mode within the cavity corresponds to a quantum harmonic oscillator with equally spaced energy levels, as depicted in Fig.~\ref{fig1}c. Each level corresponds to a Fock state $\ket{n}$ with $n$ photons ($n \in \{0,1,2\dots \}$). Leveraging the high-dimensional Hilbert space of the cavity, we encode the 02 erasure qubit using the Fock-state logical codewords:
\begin{eqnarray}
\ket{0_\mathrm{L}} = \ket{0}, \quad \ket{1_\mathrm{L}} = \ket{2}.
\end{eqnarray}  
In Fig.~\ref{fig1}d, we present the measured phase-space Wigner functions of four cardinal states on the logical Bloch sphere within the code space. For this erasure qubit, the dominant single-photon loss would bring the logical states into a leakage state $\ket{E}=\ket{1}$, which can be detected and flagged as a dominant erasure state. We measure an erasure rate from $\ket{1_\mathrm{L}}$ to $\ket{E}$ of $(0.24~\mathrm{ms})^{-1}$ due to dominant single-photon loss, substantially larger than the erasure rate from $\ket{0_\mathrm{L}}$ to $\ket{E}$ of $(64.7~\mathrm{ms})^{-1}$ due to thermal excitation (see Supplementary Sec.~III), yielding an erasure bias ratio over 265. This biased-erasure qubit provides more noise structures than conventional erasure qubits, since detecting an erasure $\ket{E}$ implies the qubit was initially in $\ket{1_\mathrm{L}}$, thereby revealing additional prior state information.

A general quantum circuit for the 02 erasure qubit is illustrated in Fig.~\ref{fig1}e, which includes state preparation (green block), logical operation (orange block), logical measurement (red block), and state classification (grey block). State preparation is performed via quantum optimal control (QOC) pulses~\cite{Heeres2017}, followed by additional ancilla and cavity check measurements to verify the state preparation and improve the preparation fidelity by flagging failures with postselection. Logical operations $U_\mathrm{L}$, also implemented with QOC pulses, are interleaved with mid-circuit erasure detections to mitigate leakage errors during gate operations. This mid-circuit detection relies on quantum non-demolition measurements of the photon-number parity~\cite{sun2014} that distinguishes the odd-photon-number erasure state from even-photon-number logical states. Final end-of-line logical measurement and state classification discriminate among $\ket{0_\mathrm{L}}$, $\ket{1_\mathrm{L}}$, and $\ket{E}$ based on all prior check outcomes. An erasure is flagged if any check returns `1'; only when all checks return \{`0'\} is the final measurement accepted as valid. 

\begin{figure*}[t]
    \centering    
    \includegraphics{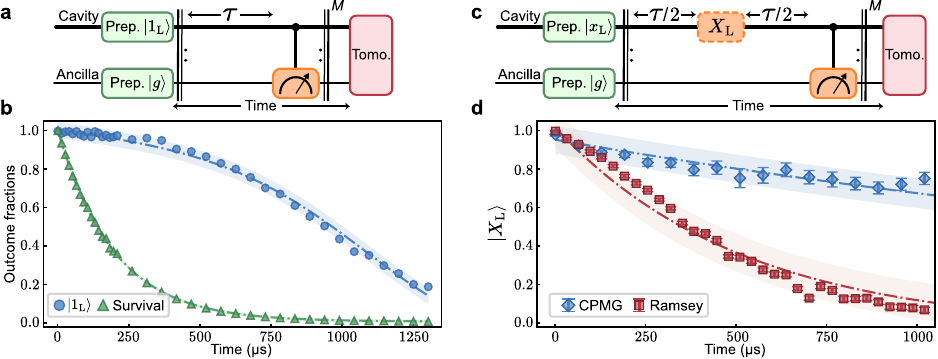}
    \caption{\textbf{Measurement of logical relaxation and dephasing rates for the 02 erasure qubit.}
\textbf{a.} Experimental sequence for measuring the logical relaxation rate. An initial $\ket{1_\mathrm{L}}$ undergoes $M$ rounds of erasure detections at a fixed repetitive interval $\tau=11.9$~\textmu s. 
\textbf{b.} Measured postselected population in $\ket{1_\mathrm{L}}$ (blue circles) and survival probability (green triangles) as a function of the total evolution time. The fitting to the population in $\ket{1_\mathrm{L}}$ yields a residual relaxation rate of $(8.9 \pm 0.5~\mathrm{ms})^{-1}$. The survival probability is exponentially fitted, giving an erasure rate of $(197 \pm 0.2~\text{\textmu} \mathrm{s})^{-1}$. The error bars are smaller than the marker size and therefore not shown. 
\textbf{c.} Experimental sequence for characterizing logical dephasing rate: a Ramsey-type sequence (without logical $X_\mathrm{L}$) and a CPMG-type sequence (with inserted $X_\mathrm{L}$), both interleaved with repeated erasure detections. 
\textbf{d.} Measured phase coherence $\langle X_{\mathrm{L}}\rangle$ for the Ramsey (red squares) and CPMG (blue diamonds) sequences, both fitted exponentially. The extracted logical dephasing rates are $(3.1 \pm 0.2~\mathrm{ms})^{-1}$ and $(0.52 \pm 0.07 ~\mathrm{ms})^{-1}$, respectively. The error bars represent $\pm 1 \sigma$ standard errors obtained from bootstrapping. The shaded areas denote the 95\% confidence prediction regions.
} 
\label{fig3}
\end{figure*}

\noindent \textbf{Logical measurement and erasure detection}{\large\par}
\noindent We first characterize the logical measurement of the erasure qubit using the sequence in Fig.~\ref{fig2}a. The logical measurement employs cascaded photon-number modulo parity measurements~\cite{deng2024}---$\mathrm{mod}(n,2)$ and $\mathrm{mod}(n,4)$---to simultaneously identify all four Fock states $\ket{n}$ with $n\in \{0,1,2,3\}$ based on the binary outcomes \{`00', `01', `10', `11'\} of the two parity measurements (see Methods). As illustrated by colored arrows in Fig.~\ref{fig2}b, we identify logical assignment errors, erasure identification errors, and leakage identification errors based on mismatches between prepared states and measured outcomes. As shown in Fig.~\ref{fig2}c, the average logical assignment error---defined as the probability of misidentifying $\ket{2}$ as `00' or $\ket{0}$ as `10'---is measured to be $(0.97 \pm 0.03)\%$.

We then evaluate the mid-circuit erasure detection using the experimental sequence in Fig.~\ref{fig2}a, where an erasure check is inserted between state preparation and final logical measurement. This experiment yields two quantitative performance metrics: false-positive and false-negative errors of the mid-circuit erasure detection, based on the detection outcomes conditioned on the final logical measurement. As shown in Fig.~\ref{fig2}d, false-positive error (incorrectly flagging an erasure) is determined to be $(0.22 \pm 0.02)\%$ averaged over the logical states. False-negative error (failing to detect an actual erasure) is measured to be $(0.69 \pm 0.03)\%$ on average, indicating that more than 99.3\% of leakage events are successfully converted into erasures.

Finally, we quantify the dephasing error induced by each erasure detection using an echoed experiment, with the sequence shown in Fig.~\ref{fig2}e. In this experiment, $M$ erasure checks are evenly inserted over a fixed total evolution duration $\tau_\mathrm{tot}=150$~\textmu s. From final Wigner tomography measurements, we extract the phase coherence $\langle X_{\mathrm{L}}\rangle$ of the erasure logical qubit as a function of $M$, as shown in Fig.~\ref{fig2}f. Here $X_\mathrm{L}=\ket{0_\mathrm{L}}\bra{1_\mathrm{L}}+\ket{1_\mathrm{L}}\bra{0_\mathrm{L}}$ represents a logical Pauli-$X$ operator on the erasure qubit. The results reveal that the phase coherence initially improves due to the correct elimination of erasure errors, and then decays linearly with $M$ due to the accumulated dephasing induced by erasure checks. From this decay, we extract an induced dephasing error of $0.26\%$ per erasure detection.

To evaluate these error rates in the context of surface code concatenation, we benchmark against target thresholds of 0.5\% erasure error and 0.5\% Pauli error, which are well within the correctable threshold for standard erasure qubits~\cite{kubica2023}. The average false-positive rate of $0.22\%$ is directly compared to the target 0.5\% erasure error, while the missed-erasure probability of 0.004\% derived from the false-negative rate (see Methods) and erasure-detection-induced dephasing error of 0.26\% are compared to the target 0.5\% Pauli error~\cite{levine2024}. Moreover, these benchmarks are for standard erasure qubits; for biased-erasure qubits, the effective thresholds are expected to be even more relaxed due to the structured erasure channel~\cite{sahay2023}.

\noindent \textbf{Characterizing logical errors during idling and gates}{\large\par}
\noindent In addition to efficient erasure detection and logical measurement capabilities, an erasure qubit should exhibit a strong hierarchy between erasure errors and residual errors during both idling and gate operations within the logical subspace.

We first quantify the logical relaxation error rate from $\ket{1_\mathrm{L}}$ to $\ket{0_\mathrm{L}}$ of the erasure qubit during idling evolution. As shown in Fig.~\ref{fig3}a, we initially prepare the erasure qubit in logical state $\ket{1_\mathrm{L}}$, then apply $M$ rounds of mid-circuit erasure detections at a fixed repetitive interval $\tau=11.9$~\textmu s (with a single cycle duration $T=13.0$~\textmu s), and finally perform a logical measurement via Wigner tomography to reconstruct the logical state. The detection interval $\tau$ is chosen to balance multiphoton-loss-induced logical relaxation against detection-induced errors (see Methods). Figure~\ref{fig3}b shows the measured survival probability (green) as a function of the total evolution time, from which we extract an erasure rate of $\gamma_\mathrm{erasure} = (0.2~\mathrm{ms})^{-1}$ via an exponential fitting. The quantum dynamics of the population in logical state $\ket{1_\mathrm{L}}$ can be modeled via the analytical expression:
\begin{eqnarray}
P_{\ket{1_\mathrm{L}}} = \frac{e^{\kappa_c\tau}+1}{(e^{\kappa_c\tau}-1)e^{2\kappa_c\tau t/T} + 2},
\label{eq:P1L_main}
\end{eqnarray}
derived from completely positive trace-preserving maps of cavity decay and erasure detection channels (see Methods). 
At short times ($\sim 150$~\textmu s), $P_{\ket{1_\mathrm{L}}}$ decays linearly with an intrinsic logical relaxation rate $\gamma_\mathrm{int}=(\kappa_c\tau)^2/T$ as postselection against erasures effectively preserves logical information. As shown in Fig.~\ref{fig3}b, the measured $P_{\ket{1_\mathrm{L}}}$ (blue circles) is fitted to the above equation with an additional exponential term to account for the residual relaxation rate $\gamma_\mathrm{res}$ of the erasure detection (see Methods and Supplementary Sec.~IV-B). The fitting result (blue dash-dotted line) yields $\gamma_\mathrm{res}=(8.9 \pm 0.5~\mathrm{ms})^{-1}$, corresponding to an effective logical relaxation rate of $\gamma_\mathrm{int}+\gamma_\mathrm{res}=(6.2\pm0.2~\mathrm{ms})^{-1}$. The ratio between erasure rate and effective logical relaxation rate thus exceeds $(0.2~\mathrm{ms})^{-1}/(6.2~\mathrm{ms})^{-1} = 31.5 \pm 1.0$

We next characterize the logical dephasing rate of the 02 erasure qubit using a Ramsey-type sequence (Fig.~\ref{fig3}c). The measured phase coherence (red squares) of the logical qubit after postselection against erasures is presented in Fig.~\ref{fig3}d. However, this biased-erasure qubit is susceptible to no-jump backaction~\cite{michael2016}, described by the Kraus operator $E_0 = e^{-\frac{\kappa_\mathrm{c}t}{2} a^\dagger a }$, where $a$ and $a^\dagger$ are cavity ladder operators. This no-jump backaction reduces the relative amplitude of $\ket{1_\mathrm{L}}$ to $\ket{0_\mathrm{L}}$, thus degrading the phase coherence of the erasure qubit. To mitigate this, we insert a logical $X_\mathrm{L}$ gate (implemented using QOC pulses) at the midpoint of the idling evolution time $\tau$. The no-jump evolution of an initial logical superposition state $\ket{x_\mathrm{L}}$ can be expressed as (ignoring the normalization factors)
\begin{eqnarray}
\ket{0_\mathrm{L}} + \ket{1_\mathrm{L}} &\xrightarrow{\mathrm{waiting}\, \tau/2}& \ket{0_\mathrm{L}} + e^{-\kappa_\mathrm{c}\tau/2}\ket{1_\mathrm{L}} \notag\\
&\xrightarrow{\mathrm{applying}\, X_\mathrm{L}}& \ket{1_\mathrm{L}} + e^{-\kappa_\mathrm{c}\tau/2}\ket{0_\mathrm{L}} \\
&\xrightarrow{\mathrm{waiting}\, \tau/2}& \ket{1_\mathrm{L}} + \ket{0_\mathrm{L}},\notag
\end{eqnarray}
effectively echoing out the no-jump effect and preserving the phase coherence of the erasure qubit. Repeating this echoed sequence over $M$ rounds with interleaved erasure detections resembles a Carr-Purcell-Meiboom-Gill (CPMG) sequence~\cite{Meiboom1958}, as shown in Fig.~\ref{fig3}c, yielding a slower decay in $\langle X_\mathrm{L} \rangle$ after postselection against erasures. An exponential fit (blue dash-dotted line) to the CPMG result (blue diamonds) yields a logical dephasing rate of $(3.1 \pm 0.2~\mathrm{ms})^{-1}$, more than 15 times lower than the erasure rate.

\begin{figure}[t]
    \includegraphics{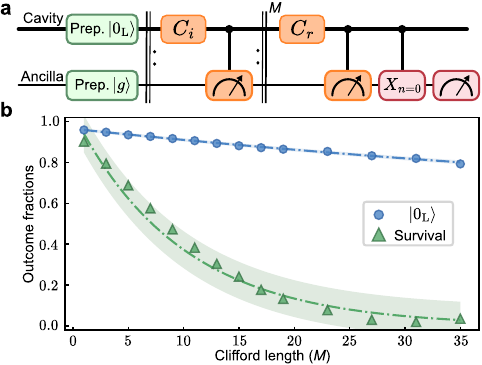}
    \caption{\textbf{Characterization of logical gate errors via randomized benchmarking.} 
    \textbf{a.} Quantum circuit for randomized benchmarking. Each Clifford gate is followed by a mid-circuit erasure detection to suppress leakage errors during gate operation. A final selective $\pi$-pulse is applied to measure the population in $\ket{0_\mathrm{L}}$. 
    \textbf{b.} Measured survival probability (green triangles) and population in $\ket{0_\mathrm{L}}$ (blue circles) as a function of Clifford length $M$. An exponential fit to the survival probability gives an erasure rate of $(4.50 \pm 0.05) \times 10^{-2}$ per gate. The $\ket{0_\mathrm{L}}$ state population is fitted to an exponential decay function with an offset of 0.5, resulting in a residual error of $ (2.86 \pm 0.07) \times 10^{-3}$ per gate. Error bars are smaller than the markers and thus are not shown. The shaded areas represent the 95\% confidence prediction regions. 
    }
    \label{fig4}
\end{figure}

The postselected relaxation and dephasing rates of the erasure qubit not only demonstrate a strong error hierarchy relative to the erasure rate during idling, but also allow extraction of the average channel fidelity decay rate as $\Gamma_{\{02\}}^\mathrm{erasure}=[(6.2~\mathrm{ms})^{-1} + 2\times(3.1~\mathrm{ms})^{-1} ]/3=(3.7 \pm 0.2~\mathrm{ms})^{-1}$ assuming amplitude damping and white noise dephasing channels in the qubit~\cite{sivak2023}. For comparison, the best physical qubit encoded using the two-lowest Fock states \{$\ket{0}$, $\ket{1}$\} achieves $\Gamma_{\{01\}}=(1/T_\mathrm{1,c} + 2/T_\mathrm{2R,c})/3\approx(0.62~\mathrm{ms})^{-1}$, defining the break-even point in this system. As a result, the postselected decoherence rates of the erasure qubit exhibit a gain factor of $G=\Gamma_{\{01\}}/\Gamma_{\{02\}}^\mathrm{erasure}=(6.0\pm0.3)$ beyond break-even with quantum error detection. Going further, larger enhancement of the decoherence rates without postselection can be achieved by implementing full QEC with both erasure detection and correction via logical qubit reset~\cite{li2024} and code concatenation~\cite{sahay2023} of the erasure qubits. 

Finally, we validate the error hierarchy during gate operations using Clifford randomized benchmarking~\cite{magesan2011scalable}. Single-qubit Clifford gates within the logical codewords are constructed using a logical $X_{\mathrm{L}}/2$ gate via numerically optimized pulses and virtual phase rotations (Supplementary Sec.~V). In the experimental sequence shown in Fig.~\ref{fig4}a, we insert a mid-circuit erasure detection after each Clifford gate to capture leakage into the error subspace. As shown in Fig.~\ref{fig4}b, the measured survival probability in the logical subspace as a function of Clifford length $M$ indicates an erasure rate of $(4.50 \pm 0.05)\times 10^{-2}$ per gate, which is higher than the idling erasure rate due to additional leakage from control pulses. After postselection against erasures, the measured state probability of $\ket{0_\mathrm{L}}$ is exponentially fitted, giving a residual error of $(2.86 \pm 0.07) \times 10^{-3}$ per gate, approximately 16 times smaller than the erasure error during gate. 

\smallskip{}
\noindent \textbf{\large{}Conclusions}{\large\par}
\noindent We have experimentally realized a biased-erasure qubit encoded in the vacuum and two-photon Fock states of a single 3D superconducting microwave cavity. This qubit exhibits a large erasure bias ratio of about 265, where erasure error from one logical basis state dominates over that from the other one. Using a single transmon ancilla, we implement hardware-efficient logical measurements and mid-circuit erasure detections, achieving a logical assignment error below 1\% and successfully converting over 99.3\% of leakage events into detected erasures. Using postselection against erasures, we demonstrate that the erasure qubit exhibits a strong error hierarchy between erasure errors and logical errors during both idling and gate operations within the logical subspace. The postselected erasure qubit during idling indicates a coherence gain of about 6.0 beyond the break-even point. Further improvements could be achieved by utilizing the third level of the transmon ancilla for fault-tolerant parity measurements~\cite{Rosenblum2018} and exploiting higher-order 0N encodings to convert multi-photon loss events into detectable erasures.

The demonstrated biased-erasure qubit offers a key building block for integrating into higher-level stabilizer codes, which can exploit both its erasure and bias properties to achieve higher thresholds with reduced resource overheads~\cite{sahay2023}. A critical next step will be to develop entangling operations between biased-erasure qubits, for example, through ancilla-mediated selective photon-number-dependent phase gates~\cite{heeres2015PRL,you2025PRA} or parametric sideband drives~\cite{s2015PRA, huang2025fast}. Combining these entangling gates with the already demonstrated capabilities---logical measurements, mid-circuit erasure detections, and single-qubit logical gates---would constitute the full ingredients required for error-detected biased-erasure qubits towards concatenations into a large scale~\cite{sahay2023}. Apart from quantum computation, the 0N erasure qubit could also promise potential applications in quantum-enhanced metrology~\cite{Wang2019NC, McCormick2019}, offering improved sensitivity with erasure detection~\cite{niroula2023}.

\textit{Note added}--After finalizing the experiment, we became aware of a similar demonstration of biased-erasure encoding in a superconducting cavity~\cite{koottandavida2025biased}.

\vbox{}

%

\clearpage{}
\setcounter{figure}{0} 
\noindent \textbf{\large{}Methods}{\large\par}

\noindent \textbf{Mid-circuit erasure detection}

\noindent For the 02 erasure qubit, the logical subspace contains only even-photon-number states, while erasure states exhibit odd-photon-number parity. Mid-circuit erasure detection is therefore implemented by performing a quantum nondemolition measurement of the photon-number parity $e^{i\pi a^\dagger a}$, which serves as an error syndrome that distinguishes erasure states from logical states. In our experiment, the parity measurement is achieved by inserting a controlled-phase gate $C_\pi = \ket{g}\bra{g} \otimes I + \ket{e}\bra{e} \otimes e^{i \pi a^\dagger a}$ between $X/2$ and $-X/2$ rotations on the transmon ancilla. Here $\ket{g}$ and $\ket{e}$ denote the ground and excited states of the ancilla, and $C_\pi$ is realized via the dispersive coupling between the cavity and the ancilla~\cite{sun2014}. Subsequent ancilla measurement reveals the photon-number parity of the cavity states: measuring the ancilla in $\ket{g}$ (outcome `0') indicates the erasure qubit remains in the code space, while $\ket{e}$ (outcome `1') indicates an odd-parity state, thereby flagging an erasure.

\vbox{}
\noindent \textbf{Cascaded photon-number modulo parity measurements}

\noindent The end-of-line logical measurement of the erasure qubit is implemented using cascaded photon-number modulo parity measurements~\cite{deng2024}---$\mathrm{mod}(n,2)$ and $\mathrm{mod}(n,4)$---which together resolve all four Fock states $\ket{0}$, $\ket{1}$, $\ket{2}$, and $\ket{3}$. The first measurement, $\mathrm{mod}(n,2)$, corresponds to the conventional photon-number parity measurement that projects the cavity onto even or odd photon-number subspace depending on the ancilla measurement outcome $b_1=0$ or 1 as described above. The second measurement, $\mathrm{mod}(n,4)$, is performed by employing a controlled-phase gate $C_{\pi/2}$ between ancilla rotations conditionally on the previous outcome $b_1$. The combined two-bit outcome `$b_2 b_1$' of the $\mathrm{mod}(n,4)$ and $\mathrm{mod}(n,2)$ parity measurements thus uniquely identifies the cavity Fock state: `00'$\rightarrow \ket{0}$, `01'$\rightarrow \ket{1}$, `10'$\rightarrow \ket{2}$, and `11'$\rightarrow \ket{3}$, respectively. Further details of the experimental sequence and characterization are provided in Supplementary Sec. II-B.

\vbox{}

\noindent \textbf{Missed-erasure probability}

\noindent Following the analysis in Ref.~\cite {levine2024}, the missed-erasure probability quantifies undetected erasure events that may persist and propagate into Pauli errors within the context of surface code concatenation. This probability is estimated as the product of (i) the probability of an erasure occurring within a single erasure check, $\gamma_\mathrm{erasure}T_\mathrm{check}\sim 0.55\%$, and (ii) the false negative rate of the mid-circuit erasure detection, 0.69\%. Here $T_\mathrm{check}=T-\tau$ denotes the duration of each erasure check. The resulting missed-erasure probability is therefore 0.004\%, well below the 0.5\% target Pauli error for surface-code concatenation with typical erasure qubits.

\vbox{}

\noindent \textbf{Optimal repetitive interval for erasure detection}

\noindent The repetitive interval $\tau$ between successive mid-circuit erasure detections involves a key trade-off: longer intervals increase the probability of undetectable two-photon-loss events, whereas shorter intervals lead to more accumulated detection-induced errors. To determine the optimal $\tau$, we perform logical relaxation measurements (Supplementary Fig.~S7) and logical dephasing experiments (Fig.~\ref{fig2}e) while varying the repetitive intervals of the interleaved erasure checks. Both experiments consistently indicate an optimal interval of $\tau\sim11.9$~\textmu s, which is consequently adopted in all subsequent experiments.

\vbox{}

\noindent \textbf{Analytical model for logical relaxation}

\noindent The logical relaxation dynamics of the erasure qubit are quantified by tracking the postselected population in $\ket{1_\mathrm{L}}$ versus the total evolution time $t=MT$, where $M$ is the number of experimental cycles and $T$ is the duration of a single cycle. Each cycle comprises two steps: (i) an idling for a time $\tau$ under the dominant photon-loss error channel, and (ii) a mid-circuit erasure detection that projects the erasure qubit into the logical code space. 

The free idling evolution of the cavity state $\rho$ (truncated to the lowest three Fock levels) is governed by the Lindblad master equation with dominant photon loss: $d\rho/dt = \kappa_\mathrm{c} (2a\rho a^\dagger - a^\dagger a\rho -\rho a^\dagger a)/2$. Over a finite interval $\tau$, this evolution is described by a completely positive and trace-preserving map $E_\mathrm{FE}(\rho) = \sum_{k=0}^{2}E_k\rho E_k^\dagger$~\cite{michael2016}, with the Kraus operators $E_k=\sqrt{\frac{(1-e^{-\kappa_\mathrm{c}\tau})^k}{k!}}e^{-\frac{\kappa_\mathrm{c}\tau a^\dagger a}{2} }a^k$. Here, $k=0$ corresponds to the no-jump backaction, while $k>0$ describes $k$-photon loss events. 
The mid-circuit erasure detection projects the cavity state onto the logical code space, effectively removing any population that leaked into $\ket{1}$. This process is modeled by the quantum channel: $\mathcal{E}_{\mathrm{ED}}(\rho) = \frac{\mathcal{P}_\mathrm{C} \rho \mathcal{P}_\mathrm{C}^\dagger}{\operatorname{Tr}(\mathcal{P}_\mathrm{C} \rho \mathcal{P}_\mathrm{C}^\dagger)}$, where $\mathcal{P}_\mathrm{C}=\ket{0_\mathrm{L}}\bra{0_\mathrm{L}} + \ket{1_\mathrm{L}}\bra{1_\mathrm{L}}$ represents the projector onto the code space. 
These two quantum channels together compose a full cycle: $\mathcal{E}_{\mathrm{cycle}} = \mathcal{E}_\mathrm{ED} \circ \mathcal{E}_\mathrm{FE}$. Starting from an initial state $\rho_0=\ket{2}\bra{2}$, the cavity state after $M$ cycles is $\mathcal{E}_\mathrm{cycle}^{\circ N}(\rho_0)$. The corresponding population remaining in $\ket{1_\mathrm{L}}$ is obtained analytically as $P_{\ket{1_\mathrm{L}}}=\bra{2} \mathcal{E}_\mathrm{cycle}^{\circ N}(\rho_0) \ket{2}= \frac{e^{\kappa_\mathrm{c}\tau}+1}{(e^{\kappa_\mathrm{c}\tau}-1)e^{2M\kappa_\mathrm{c}\tau}+2}$ with $M=t/T$. In the short-time limit, the population decays linearly as $P_{\ket{1_\mathrm{L}}} \approx 1 - \gamma_\mathrm{int}t$ with an intrinsic logical relaxation rate $\gamma_\mathrm{int} = (\kappa_\mathrm{c}\tau)^2/T=(20~\mathrm{ms})^{-1}$. At longer times, the decay becomes exponential, $P_{\ket{1_\mathrm{L}}}\propto e^{-2\kappa_\mathrm{c}\tau t/T}$, dominated by the increasing probability of consecutive single-photon loss events. Further details can be found in Supplementary Sec.~IV-B. 

To account for residual relaxation due to experimental imperfections (e.g., ancilla errors, cavity thermal excitations, detection-induced errors), the measured $P_{\ket{1_\mathrm{L}}}$ in Fig.~\ref{fig3}b is fitted to an effective model that introduces an additional exponential decay, $P_{\ket{1_\mathrm{L}},\mathrm{eff}}(t) = A\, P_{\ket{1_\mathrm{L}}}(t)\, e^{-\gamma_{\mathrm{res}} t} + B$. Here, $A$ and $B$ are fitting parameters to account for state-preparation and readout imperfections, and $\gamma_\mathrm{res}$ is the residual relaxation rate. Expanding this expression at short times gives a linear decay $\sim 1-(\gamma_\mathrm{int} + \gamma_\mathrm{res})t$, yielding the total effective logical relaxation rate $\gamma_\mathrm{tot} = \gamma_\mathrm{int} + \gamma_\mathrm{res}$.

\vbox{}

\smallskip{}

\noindent \textbf{\large{}Data availability}{\large\par}

\noindent Source data are provided with this paper. Further data relevant to this study are available from the corresponding authors upon request.

\smallskip{}

\noindent \textbf{\large{}Code availability}{\large\par}

\noindent The code used in this study is available from the corresponding authors upon reasonable request.

\smallskip{}

\noindent \textbf{\large{}Acknowledgment}{\large\par}

\noindent This work was supported by the National Natural Science Foundation of China (Grants No.~12422416, No.~12274198, No.~12575025), the Quantum Science and Technology-National Science and Technology Major Project (Grants No.~2024ZD0302300, No.~2021ZD0301703), and the Guangdong Basic and Applied Basic Research Foundation (Grant No.~2024B1515020013).

\smallskip{}

\noindent \textbf{\large{}Author contributions}{\large\par}

\noindent Y.X. conceived the idea and supervised the project. J.M., Q.L., and X.D. performed the measurements, analysed the data, and conducted the numerical simulations under the supervision of Y.X. L.Z. and S.L. provided support in the fabrication of the transmon ancilla. Y.C., Z.N., L.H., and P.Z. contributed to the experimental discussion. J.M., Q.L., X.D., and Y.X. wrote the manuscript with feedback from all authors.

\smallskip{}

\noindent \textbf{\large{}Competing interests}{\large\par}

\noindent The authors declare no competing interests.

\clearpage{}

\end{document}


\title{Supplementary Information for “A biased-erasure cavity qubit with hardware-efficient quantum error detection”}

\author{Jiasheng Mai}
\thanks{These authors contributed equally to this work.}
\affiliation{Southern University of Science and Technology, Shenzhen 518055, China}
\affiliation{International Quantum Academy, Shenzhen 518048, China}

\author{Qiyu Liu}
\thanks{These authors contributed equally to this work.}
\affiliation{International Quantum Academy, Shenzhen 518048, China}

\author{Xiaowei Deng}
\thanks{These authors contributed equally to this work.}
\email{dengxiaowei@iqasz.cn}
\affiliation{International Quantum Academy, Shenzhen 518048, China}

\author{Yanyan Cai}
\affiliation{Southern University of Science and Technology, Shenzhen 518055, China}
\affiliation{International Quantum Academy, Shenzhen 518048, China}

\author{Zhongchu Ni}
\affiliation{International Quantum Academy, Shenzhen 518048, China}

\author{Libo Zhang}
\affiliation{Southern University of Science and Technology, Shenzhen 518055, China}
\affiliation{International Quantum Academy, Shenzhen 518048, China}

\author{Ling Hu}
\affiliation{International Quantum Academy, Shenzhen 518048, China}
\affiliation{Shenzhen Branch, Hefei National Laboratory, Shenzhen 518048, China}

\author{Pan Zheng}
\affiliation{International Quantum Academy, Shenzhen 518048, China}

\author{Song Liu}

\author{Yuan Xu}
\email{xuyuan@iqasz.cn}
\affiliation{International Quantum Academy, Shenzhen 518048, China}
\affiliation{Shenzhen Branch, Hefei National Laboratory, Shenzhen 518048, China}
\author{Dapeng Yu}
\affiliation{International Quantum Academy, Shenzhen 518048, China}
\affiliation{Shenzhen Branch, Hefei National Laboratory, Shenzhen 518048, China}

\maketitle

\tableofcontents

\clearpage

\section{Device and Hamiltonian}

\subsection{Experimental device and setup}
\begin{figure}[b]
	\includegraphics{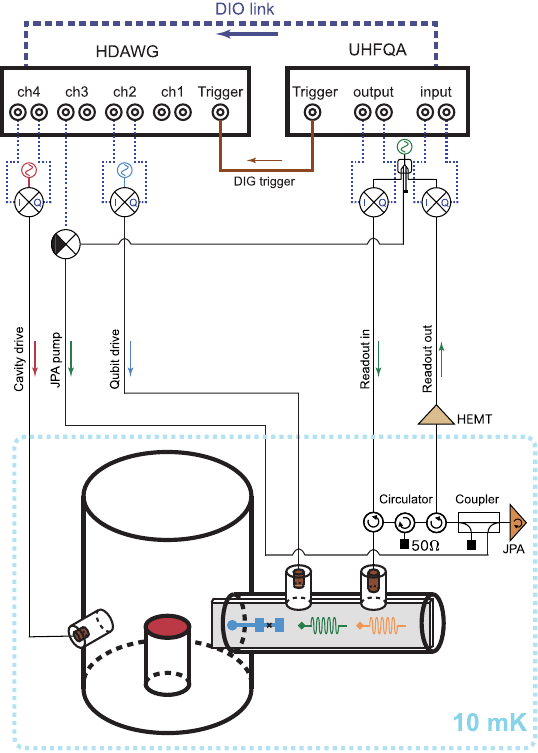}
	\caption{Measurement setup and experimental device schematic.}
	\label{figS1}
\end{figure}

The three-dimensional (3D) superconducting bosonic system used in this work shares a similar architecture with that in Ref.~\cite{ni2023,deng2024}, and the device schematic is shown in Fig.~\ref{figS1}. It comprises three components: a coaxial cylindrical cavity~\cite{reagor2016} acting as a quarter-wavelength ($\lambda/4$) transmission-line resonator, a superconducting transmon ancilla~\cite{koch2007}, and a Purcell-filtered stripline readout resonator~\cite{axline2016}. 

The coaxial cavity (depicted in red in Fig.~\ref{figS1}) is machined from high-purity (5N5) aluminum and used to encode the erasure qubit. The Josephson junction and two antenna pads of the transmon, as well as the metal striplines of the readout resonator, are fabricated lithographically on a single sapphire substrate. One of the transmon's antenna pads directly couples to the cavity mode, the other to the readout resonator. The quasi-planar Purcell-filtered readout resonator contains two $\lambda/2$ striplines: the first (green) is strongly coupled to the transmon ancilla and enables fast ancilla readout; the second (orange) is coupled to the external environment and acts as a bandpass Purcell filter, enhancing the coherence lifetimes of the transmon-cavity system.

The experimental device is installed inside a magnetic shield in a dilution refrigerator operating below $10\mathrm{mK}$. All microwave control and readout signals are generated with room-temperature electronics, with the experimental circuitry shown in Fig.~\ref{figS1}. Drive pulses for the ancilla and cavity, as well as the readout pulse, are generated by single-sideband in-phase and quadrature (IQ) modulation using waveforms from an arbitrary waveform generator (Zurich Instruments, HDAWG) and a quantum analyzer (Zurich Instruments, UHFQA). These signals are up-converted and sent to the device through coaxial cables equipped with microwave isolators, filters, and attenuators to suppress reflection and radiation noise. The reflected readout signal from the readout resonator is amplified through a multistage amplification chain: a quantum-limited Josephson parametric amplifier (JPA) at the base temperature, a high-electron-mobility-transistor (HEMT) amplifier at the $4\mathrm{K}$ stage, and a commercial low-noise RF amplifier at room temperature. The amplified readout signal is finally down-converted via IQ demodulation using the same local oscillator as employed in the readout pulse generation, then digitized and recorded by the analog-to-digital converter (ADC) of the UHFQA. 
Timing synchronization between the UHFQA and HDAWG is ensured via a digital trigger (DIG), which aligns the transmon/cavity drive pulses with the readout pulse. For fast feedback control, a digital input/output (DIO) link connects the UHFQA to the HDAWG, achieving a feedback latency of $600~\mathrm{ns}$.

\subsection{System Hamiltonian and parameters}
The 3D circuit quantum electrodynamics (QED) system comprises an anharmonic bosonic mode (the transmon ancilla) and two nearly linear bosonic modes (the cavity and readout resonator). In the dispersive regime, its effective Hamiltonian can be described as:
\begin{align}\label{H_sym}
	\begin{split}
		H/\hbar
		 & =\omega_\mathrm{q}a_\mathrm{q}^\dagger a_\mathrm{q}+\omega_\mathrm{c}a_\mathrm{c}^\dagger a_\mathrm{c}+\omega_\mathrm{r}a_\mathrm{r}^\dagger a_\mathrm{r}                                      \\
		 & -\begin{array}{c}\frac{K_\mathrm{q}}{2}a_\mathrm{q}^{\dagger2}a_\mathrm{q}^2-\frac{K_\mathrm{c}}{2}a_\mathrm{c}^{\dagger2}a_\mathrm{c}^2\end{array}                                       \\
		 & -\chi_{\mathrm{qc}}a_{\mathrm{q}}^{\dagger}a_{\mathrm{q}}a_{\mathrm{c}}^{\dagger}a_{\mathrm{c}}-\chi_{\mathrm{qr}}a_{\mathrm{q}}^{\dagger}a_{\mathrm{q}}a_{\mathrm{r}}^{\dagger}a_{\mathrm{r}}
	\end{split}
\end{align}
where $\omega_\mathrm{q,c,r}$ denote the frequencies of the transmon, the cavity and the readout resonator, respectively; $a_\mathrm{q,c,r}$ and $a^\dagger_\mathrm{q,c,r}$ are their corresponding annihilation and creation operators; $K_\mathrm{q,c}$ represent the self-Kerrs of the transmon and the cavity mode; and $\chi_\mathrm{qc}$ and $\chi_\mathrm{qr}$ are the cross-Kerrs between the transmon ancilla and the cavity, and between the transmon and the readout resonator, respectively. All these parameters are experimentally characterized and summarized in Table~\ref{tab:parameter_sum}. Additionally, parameters not explicitly included in the Hamiltonian, such as the second-order cross-Kerr interaction between the transmon ancilla and the cavity, are also listed.

The coherence times of the three modes in this system are also listed in Table~\ref{tab:parameter_sum}. The transmon exhibits a longitudinal relaxation time $T_\mathrm{1, q}\simeq 141$~\textmu s and a Ramsey decay time $T_\mathrm{2R,q} \simeq 117$~\textmu s, yielding a pure dephasing time $T_\mathrm{\phi,q} \simeq 200$~\textmu s. The bosonic cavity mode shows a single-photon lifetime $T_\mathrm{1, c} \simeq 466$~\textmu s and a Ramsey time $T_\mathrm{2R,c} \simeq 735$~\textmu s, corresponding to a pure dephasing time $T_\mathrm{\phi, c} \simeq 3073$~\textmu s. The readout resonator has a linewidth of $\kappa_\mathrm{r}/2\pi= 1.542$~MHz, equivalent to a decay time of $\sim 103$~ns. The transmon readout is optimized with a pulse duration of approximately $800$~ns, achieving an average fidelity of 0.994 and an average quantum non-demolition (QND) probability of $0.990$. A delay of $\sim 800$~ns is required after each readout pulse to allow the readout resonator to relax to its ground state.

\begin{table*}[htbp]
	\begin{ruledtabular}
		\centering
		\caption{Hamiltonian parameters and coherence times}
		\label{tab:parameter_sum}
		\begin{tabular}{cccccc}
			Measured parameter                                           & Cavity &   & Transmon qubit &   & Readout resonator \\
			\cline{1-6}
			Mode frequency $\omega_{\mathrm{c,q,r}}/2\pi$                         & $6.592~\mathrm{GHz}$              &   & $5.249~\mathrm{GHz}$              &   & $8.540~\mathrm{GHz}$              \\
			Self-Kerr $K_{\mathrm{c,q,r}}/2\pi$                                   & $3.98~\mathrm{kHz}$             &   & $222.96~\mathrm{MHz}$              &   & -            \\
			Cross-Kerr $\chi_{\mathrm{qc}}/2\pi,\chi_{qr}/2\pi$                   &                & $1.69~\mathrm{MHz}$ &                & $1.01~\mathrm{MHz}$ &                \\
			Second-order Cross-Kerr $\chi_{\mathrm{qc}}^{'}/2\pi$ &                & $1.45~\mathrm{kHz}$ &                & - &                \\
			\cline{1-6}
			Relaxation time $T_{\mathrm{1}}$                                     &     
            $466 $~\textmu s  
            &   & $141 $~\textmu s              &   &  $103~\mathrm{ns}$              \\
            Decoherence time $T_2$                           & 
            $735 $~\textmu s
            &   & $117$~\textmu s             &   & -              \\			Pure dephasing time $T_\mathrm{\varphi}$                           & 
            $3073$~\textmu s
            &   & $200$~\textmu s             &   & -              \\
			Thermal population $P_{\mathrm{th}}$                                  & $0.72\%$              &   & $5.34\%$              &   & -              \\
		\end{tabular}
	\end{ruledtabular}
\end{table*}

\section{Logical state preparation and measurement}

\subsection{Logical state preparation}
\begin{figure}[htbp]
\centering
\includegraphics{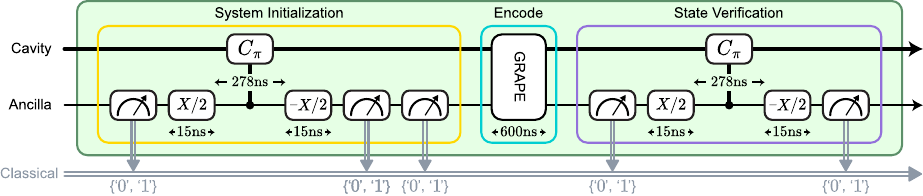}
\caption{Experimental sequence for the logical state preparation of the $02$ erasure state, which comprises three main stages: ancilla and cavity initialization, cavity state encoding using optimal control pulses, and state verification of the ancilla and cavity system.}
	\label{fig:state_prep}
\end{figure}

Arbitrary logical states of the 02 erasure qubit are prepared using the experimental sequence illustrated in Fig.~\ref{fig:state_prep}. This procedure consists of three stages: initialization, encoding, and verification. 

Initializing the transmon-cavity system (yellow block in Fig.~\ref{fig:state_prep}) is achieved by first measuring the transmon state, then performing a photon-number parity measurement of the cavity, and finally implementing another transmon measurement check. Postselecting the transmon in ground state in the three measurements would project the transmon-cavity system into the joint ground state $\ket{g,0}$, thus suppressing residual thermal excitations in both transmon and cavity in the initial states.

Once the ground state is prepared, quantum optimal control (QOC) pulses with a duration of 600~ns are applied simultaneously to the transmon and the cavity (cyan block). These pulses are numerically optimized based on the gradient ascent pulse engineering (GRAPE) technique~\cite{khaneja2005} to implement the desired state transformation that encodes the target logical states~\cite{Heeres2017} of the 02 erasure qubit. 

To ensure high fidelity, a state verification step (purple block) is performed after encoding. It consists of an ancilla measurement check followed by a cavity parity measurement check. Postselections of these checks eliminate imperfect state preparations due to decoherence and spurious excitations during the optimal control pulses.

\subsection{End-of-the-line logical measurement}
In order to extract the logical state information at the end of the experimental sequence, we implement three types of end-of-the-line logical measurements in our experiment. In this section, we provide a detailed description of the three types of logical measurements: cascaded photon-number parity measurement, Wigner tomography, and photon-number-resolved measurement.

\subsubsection{Cascaded parity measurement}

\begin{figure}[b]
\centering
\includegraphics{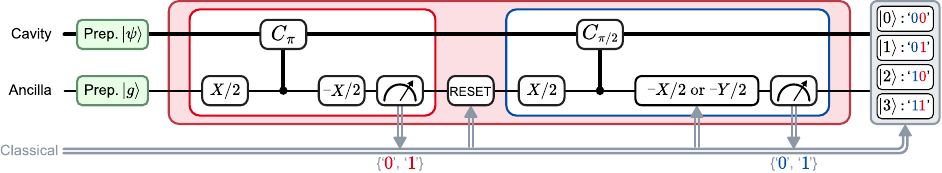}
	\caption{Experiment sequence for the cascaded parity measurement. It contains two photon-number modulo parity measurements $\mathrm{mod}(n,2)$ and $\mathrm{mod}(n,4)$, which simultaneously discriminate the cavity state among the four Fock states $\ket{0}$, $\ket{1}$, $\ket{2}$, and $\ket{3}$ from the combined outcomes of the two measurements.}
	\label{fig:mod2mod4_circuit}
\end{figure}

To identify the logical assignment errors, erasure errors, and the leakage identification errors of the 02 erasure qubit, it is necessary to fully discriminate cavity states among the four Fock states $\{\ket{0}, \ket{1}, \ket{2}, \ket{3}\}$ within the cavity. Here, we employ a cascaded measurement of the photon-number modulo parities $\mathrm{mod}(n,2)$ and $\mathrm{mod}(n,4)$ to achieve this state discrimination, with the experimental sequence shown in Fig.~\ref{fig:mod2mod4_circuit}. 

The $\mathrm{mod}(n,2)$ parity measurement is achieved by inserting a controlled-phase gate $C_{\pi} = \ket{g}\bra{g} \otimes I + \ket{e}\bra{e} \otimes e^{i \pi a^\dagger a}$ between $X/2$ and $-X/2$ rotations on the transmon ancilla. Here, $a^\dagger$ and $a$ represent the cavity ladder operators, and we ignore the subscript in the following for simplicity. This sequence transforms the initial joint cavity-ancilla state $\sum_n c_{n}\ket{n}\ket{g}$ into a final state $\sum_n[c_{n} \ket{n}(\ket{g}+i\ket{e})/2-ie^{in\pi}\ket{n}(i\ket{g}+\ket{e})/2]$, yielding an excited state population of $P_{\mathrm{e}}=\sum_{n} \frac{1-\cos{n\pi}}{2}|c_{n}|^2$. Thus, measuring the transmon ancilla in the ground ($b_1=$ `0') or excited ($b_1=$ `1') states projects the cavity onto even ($\{\ket{0}, \ket{2}\}$) or odd ($\{\ket{1}, \ket{3}\}$) photon-number subspaces, respectively. 

Following an active reset of the transmon ancilla via measurement-based feedback control, the second $\mathrm{mod}(n,4)$ parity measurement is implemented in a similar way by employing a controlled-phase gate $C_{\pi/2} = \ket{g}\bra{g} \otimes I + \ket{e}\bra{e} \otimes e^{i \frac{\pi}{2} a^\dagger a}$ between ancilla rotations conditionally on the previous measurement outcomes $b_1$. Specifically, when the ancilla is measured in ground state ($b_1=$ `0') in the first parity measurement, the $C_{\pi/2}$ gate is applied between $X/2$ and $-X/2$ gates on the ancilla, mapping the even cavity state of the joint system to $\sum_{n=0,2}[c_{n} \ket{n}(\ket{g}+i\ket{e})/2-ie^{in\pi/2}c_n\ket{n}(i\ket{g}+\ket{e})/2]$, which yields $P_{e}=\sum_{n=0,2} \frac{1-\cos{(n\pi/2)}}{2}|c_{n}|^2$. When the ancilla is measured in excited state ($b_1=$ `1') in the first parity measurement, the $C_{\pi/2}$ gate is applied between $X/2$ and $-Y/2$ rotations on the ancilla, resulting in the joint state $\sum_{n=1,3}[ic_{n} \ket{n}(\ket{g}-\ket{e})/2+ie^{in\pi/2}c_{n}\ket{n}(\ket{g}+\ket{e})/2]$ and the probability $P_{e}=\sum_{n=1,3} \frac{1-\sin{(n\pi/2)}}{2}|c_{n}|^2$.

Combining the outcomes $b_1$ and $b_2$ from the $\mathrm{mod}(n, 2)$ and $\mathrm{mod}(n, 4)$ parity measurements, the resulting two-bit string `$b_2 b_1$' uniquely identifies the cavity Fock states as follows: `00'$\rightarrow \ket{0}$, `01'$\rightarrow \ket{1}$, `10'$\rightarrow \ket{2}$, and `11'$\rightarrow \ket{3}$. Using this method, we have characterized the logical measurement and erasure detection errors in Fig. 2 in the main text.

\subsubsection{Wigner tomography}

\begin{figure}[b]
	\centering
	\includegraphics{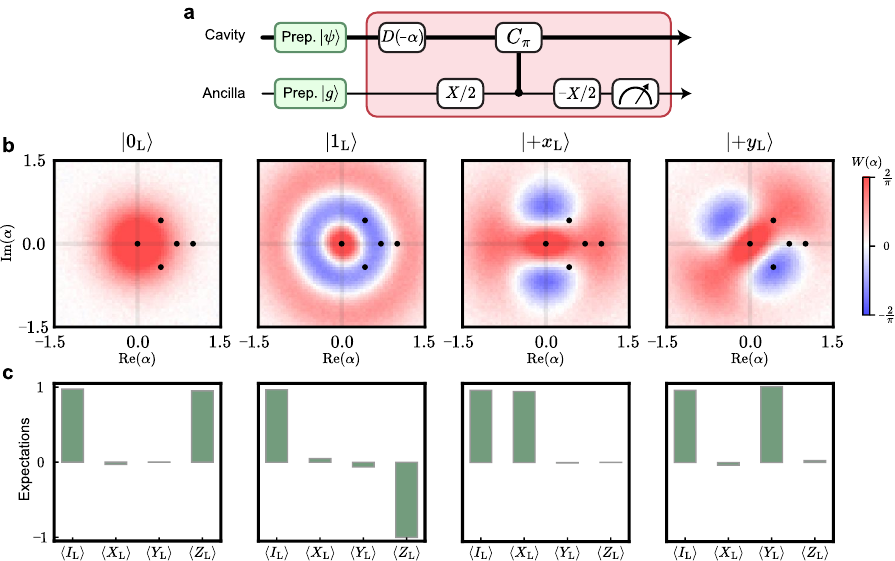}
	\caption{Wigner tomography of the erasure qubit within the logical subspace spanned by $\{\ket{0},\ket{2}\}$. \textbf{a.} Experimental sequence for the Wigner tomography measurement. \textbf{b.} Measured full Wigner functions of four logical states of the erasure qubit, along with the five Wigner points that are used for efficiently extracting the expectation values of the Pauli operators in the logical subspace. \textbf{c.} The extracted expectation values of the logical Pauli operators $\langle I_{\mathrm{L}}\rangle$, $\langle X_{\mathrm{L}}\rangle$, $\langle Y_{\mathrm{L}}\rangle$, and $\langle Z_{\mathrm{L}}\rangle$ from the five Wigner points in \textbf{b}.}
	\label{fig:fivepoints}
\end{figure}

The Wigner function of the cavity state $\rho$ can be calculated by $W(\alpha) = \frac{2}{\pi} \mathrm{Tr}[\rho P(\alpha)]$, where $P(\alpha) = D(\alpha)\Pi D(-\alpha)$ is the displaced parity operator and $\Pi = (-1)^{a^\dagger a}$ is the photon-number parity operator. As illustrated in Fig.~\ref{fig:fivepoints}a, a single value of Wigner function $W(\alpha)$ is obtained by applying a displacement $D(-\alpha)$ on the cavity, followed by a photon-number-parity measurement assisted by the ancilla.

Although the full Wigner function contains complete information about the cavity state, performing full Wigner tomography to reconstruct the state's density matrix requires sampling a sufficient number of displacement points in phase space, which is time-consuming. As we primarily focus on the logical qubit subspace spanned by Fock states $\{\ket{0}, \ket{2}\}$, the logical density matrix of the 02 erasure qubit can be reconstructed efficiently following a similar procedure in Ref.~\cite{koottandavida2024,vlastakis2015}. Specifically, only five strategically chosen displacement points are sufficient to extract the expectation values of the logical Pauli operators to reconstruct the cavity density matrix within this subspace.

The five sampling displacement points are found by projecting the displaced parity operator $P(\alpha)$ onto the logical subspace and matching the result to four logical Pauli operators $\{I_{\mathrm{L}}, X_{\mathrm{L}}, Y_{\mathrm{L}}, Z_{\mathrm{L}}\}$. Solving the equations for $\alpha$ yields 
\begin{equation}
\alpha = \{0, 1, \frac{1-i}{\sqrt{4\sqrt{2}}}, \frac{1+i}{\sqrt{4\sqrt{2}}}, \frac{1}{\sqrt{2}}\}. 
\end{equation}

This indicates that if we restrict ourselves to the logical subspace of the 02 erasure qubit, then we only need to measure these five points in phase space to reconstruct the density matrix within this subspace. Figure~\ref{fig:fivepoints}b shows the locations of these five points in phase space for each logical state. From these five displacements, we can then express the logical Pauli operators as
\begin{align}\label{Pauli_02}
	\begin{split}
		I_{\mathrm{L}} & = P(0)                                                                                                                                               \\
		X_{\mathrm{L}} & = \frac{e^2P(1)-P(0)}{2\sqrt{2}}                                                                                                                   \\
		Y_{\mathrm{L}} & = -\frac{e^{\frac{1}{\sqrt{2}}}}{2} \left[ P\left( \frac{1 - i}{\sqrt{4\sqrt{2}}} \right) - P\left( \frac{1 + i}{\sqrt{4\sqrt{2}}} \right) \right] \\
		Z_{\mathrm{L}} & = eP\left( \frac{1}{\sqrt{2}} \right) - \sqrt{2}X_{\mathrm{L}}
	\end{split}
\end{align}
After measuring the Wigner functions $W(\alpha)$ at these five points, we can directly extract the expectation values of the logical Pauli operators from Eq.~\ref{Pauli_02} for each logical state, as shown in Fig.~\ref{fig:fivepoints}c. Thus, we can directly reconstruct the density matrix $\rho$ of the cavity state within the logical subspace spanned by $\{\ket{0}, \ket{2}\}$ through the equation $\rho = \sum_{O=I,X,Y,Z}{\frac{\langle O_\mathrm{L}\rangle}{2} O_\mathrm{L} }$. 

Direct reconstruction from the linear inversion could yield unphysical density matrices due to statistical errors. We therefore employ a maximum-likelihood estimation (MLE) routine~\cite{James2001} that constrains the reconstructed state to be positive semidefinite and of unit trace, thus ensuring the reconstructed density matrix is physically valid.

To also capture leakage into the erasure state $\ket{1}$, we also construct the density matrix in the three-level Hilbert space spanned by $\{\ket{0}, \ket{1}, \ket{2}\}$ using eight sampling points:
\begin{equation}
\alpha = \left\{\,0,\; 1,\; \tfrac{1}{\sqrt{2}},\; -\tfrac{1}{\sqrt{2}},\; 
\tfrac{i}{\sqrt{2}},\; -\tfrac{i}{\sqrt{2}},\;
\tfrac{1+i}{\sqrt{4\sqrt{2}}},\;
\tfrac{1-i}{\sqrt{4\sqrt{2}}}\,\right\}.
\end{equation}
With the measured Wigner functions at these eight displacement points in phase space, we can reconstruct the cavity state that closely matches the experimental data within the qutrit subspace using a similar MLE method.

\subsubsection{Photon-number-resolved measurement}

\begin{figure}[t]
    \centering
    \includegraphics{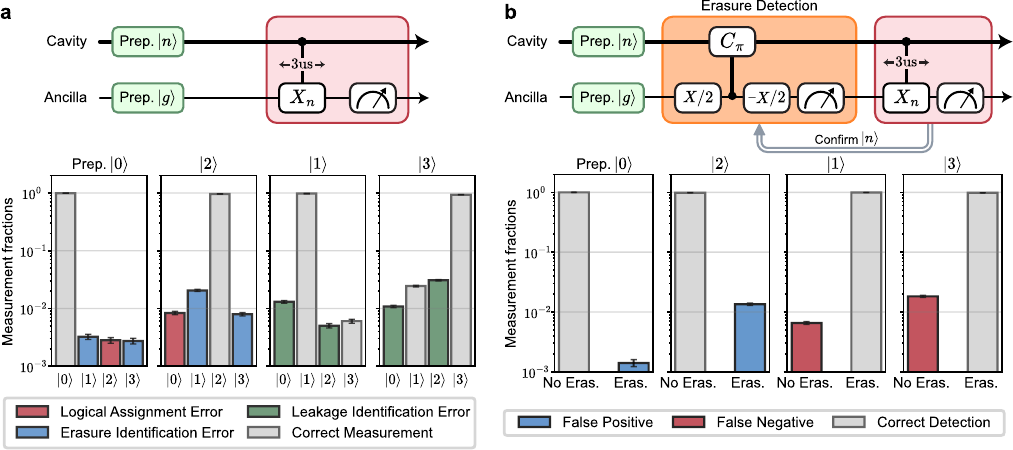}
	\caption{Characterizing photon-number-resolved logical measurement and erasure detection. 
	\textbf{a.} The experimental sequence (top) and the corresponding measurement results (bottom) for the photon-number-resolved logical measurement to extract the logical assignment errors, erasure
identification errors, and leakage identification errors.
	\textbf{b.} The logical measurement sequence with an interleaved mid-circuit erasure detection (top), resulting in the assigned erasure fractions when preparing different initial states (bottom) to extract the false-positive and false-negative rates. More than $50,000$ experimental shots were performed in these experiments.}
	\label{fig:misassignment}
\end{figure}

As an alternative approach to discriminate the cavity states among the four Fock states $\{\ket{0}, \ket{1}, \ket{2}, \ket{3}\}$, we also implement a photon-number-resolved measurement. Because the ancilla frequency depends on the photon numbers within the cavity due to the dispersive coupling~\cite{schuster2007,vlastakis2013} between them, we can directly apply a selective ancilla flip operation (with a duration of 3~\textmu s) conditioned on the cavity's specific Fock component. Measuring the ancilla state would directly reveal the photon-number populations within the cavity. 

As a supplement, we here characterize this logical measurement approach with and without interleaved mid-circuit erasure detection of the logical qubit, with the experimental sequence and results shown in Fig.~\ref{fig:misassignment}. In a similar analysis to that in Fig.~2 of the main text, we achieve an average logical assignment error of $0.6\%$, a false-positive error of $0.74\%$, and a false-negative error of $1.24\%$, which are comparable to those presented in Fig.~2 of the main text. Additionally, we use this photon-number-resolved end-of-line logical measurement to extract the recovery probability in the randomized benchmarking experiments in Fig.~4 of the main text.

\section{Characterizing logical errors during idling}
As demonstrated in the main text, the 02 erasure qubit exhibits a strong hierarchy between erasure errors and residual errors during idling within the logical subspace. In this section, we provide a detailed description of the experimental characterization. We first quantify the transition rates within the truncated Fock qutrit subspace spanned by $\{\ket{0}, \ket{1}, \ket{2}\}$ without mid-circuit erasure detection. We then detail the experimental optimization of the erasure-check interval, derive the analytical model for logical relaxation, and finally perform numerical simulations to verify the experimental results.

\subsection{Erasure errors of the logical qubit under free evolution}

\begin{figure}[t]
\includegraphics{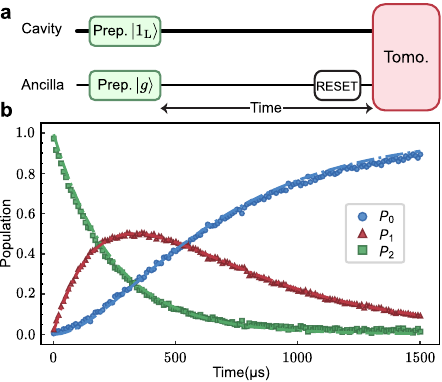}
	\caption{\textbf{a.} Experimental sequence for monitoring the free evolution of an initial logical state $|1_\mathrm{L}\rangle$.
    \textbf{b.} The measured photon-number populations $P_0$, $P_1$ and $P_2$ for Fock state $\ket{0}$, $\ket{1}$, and $\ket{2}$ as a function of the evolution time. The dash-dotted lines correspond to the global fit to the coupled rate equations to extract the transition rates.}
	\label{fig:Qutrit decay}
\end{figure}

The 02 erasure qubit in the cavity is susceptible to single-photon relaxation and excitation errors, which can cause the logical states to transition into leakage states, resulting in erasure errors. To investigate the erasure error rates, we perform an experiment to monitor the free decay evolution from an initial logical state $\ket{1_{\mathrm{L}}} = \ket{2}$ within the Fock qutrit subspace $\{\ket{0}, \ket{1}, \ket{2}\}$, with the experimental sequence shown in Fig.~\ref{fig:Qutrit decay}a. The time-dependent evolution of the Fock state populations $P_0$, $P_1$, and $P_2$ is modeled by the coupled rate equations:
\begin{equation}
    \label{eq:qutrit_ode}
    \begin{aligned}
        \frac{d P_{2}}{d t} & = -\gamma_{21} P_{2} + \gamma_{12} P_{1} \\
        \frac{d P_{1}}{d t} & = -\gamma_{10} P_{1} + \gamma_{21} P_{2} - \gamma_{12} P_{1} + \gamma_{01} P_{0} \\
        \frac{d P_{0}}{d t} & = \gamma_{10} P_{1} - \gamma_{01} P_{0}
    \end{aligned}
\end{equation}
where $\gamma_\mathrm{ij}$ denotes the transition rate from state $\ket{i}$ to $\ket{j}$.

In this experiment, an ancilla reset operation via measurement and postselection is applied before the final Wigner tomography measurement to avoid any thermal excitations of the transmon ancilla during the idling evolution. Figure~\ref{fig:Qutrit decay}b shows the measured photon-number populations $P_0$, $P_1$, and $P_2$ as a function of the evolution time, as well as a global fit to the rate equations in Eq.~\eqref{eq:qutrit_ode}. In this fit, we have constrained $\gamma_{10}=1/T_1^c$ and $\gamma_{01}=P_\mathrm{th}^c/T_1^c=(64.7~\mathrm{ms})^{-1}$, which are obtained from independent characterization experiments, yielding fitting parameters $\gamma_{21} = (244.0\pm 1.6 ~\mathrm{\mu s})^{-1}$ and $\gamma_{12}=(33.6 \pm 0.2~\mathrm{ms})^{-1}$, respectively.

\subsection{Experimental optimization of the erasure-check interval}

The interval $\tau$ between consecutive mid-circuit erasure detections is critical to characterize the logical errors of the 02 erasure qubit. An interval that is too long allows undetected two-photon-loss events to accumulate, potentially resulting in undetectable Pauli errors within the logical subspace. Conversely, an overly short interval introduces excessive operational errors due to imperfect erasure detections. Therefore, our goal is to determine an optimal detection interval $\tau$ that balances these competing effects.

\begin{figure}[tb]
\includegraphics{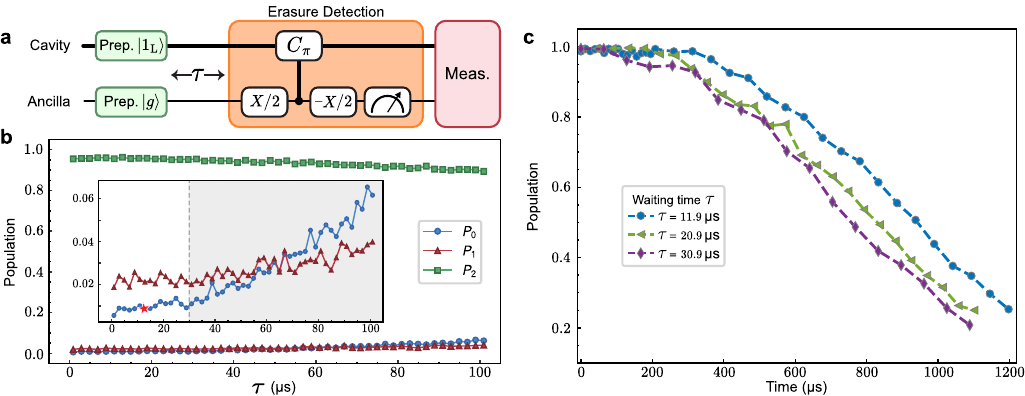}
\caption{Optimization of the repetitive interval of erasure detections.
\textbf{a.} Experimental sequence for the optimization.
\textbf{b.} Measured populations in Fock states $\ket{0}$, $\ket{1}$, and $\ket{2}$ as varying $\tau$. The shaded region ($\tau > 30$~\textmu s) in the inset indicates an apparent increase in $P_0$ due to undetected multiphoton-loss errors. The red star corresponds to $\tau \approx 11.9$~\textmu s. 
\textbf{c.} Measured population in $\ket{1_{\mathrm{L}}}$ as a function of repetitive cycles or total evolution time with three different intervals of $\tau = 11.9$~\textmu s, $20.9$~\textmu s, and $30.9$~\textmu s.}
\label{fig:Tau_ED}
\end{figure}

We first investigate the optimal $\tau$ in the logical relaxation experiment. We prepare the cavity in $\ket{2}$ and vary the interval $\tau$ between the state initialization and a mid-circuit erasure detection, with the experimental sequence shown in Fig.~\ref{fig:Tau_ED}a. The measured photon-number populations $P_0$, $P_1$, and $P_2$ are plotted as a function of $\tau$ in Fig.~\ref{fig:Tau_ED}b. The result indicates that the population in $\ket{0}$ remains at a relatively low level for $\tau < 30$~\textmu s. For longer $\tau$, the apparent increase in $P_0$ implies that undetected erasure errors have accumulated and converted into logical bit-flip errors. 

We then repeat the erasure detection $M$ cycles with different repetitive intervals $\tau$, similar to the experiment in Fig.~3b of the main text. The experimental results are shown in Fig.~\ref{fig:Tau_ED}c, which yields an optimal interval $\tau=11.9$~\textmu s in the logical relaxation experiments.

Additionally, we investigate the effect of the interval $\tau$ on the logical dephasing rate of the erasure qubit by performing an echoed experiment. In this experiment (Fig.~2e in the main text), we choose a fixed total duration and insert a variable number of erasure checks, equivalent to changing the erasure-check interval $\tau$. From the results in the main text, we obtain a consistent optimal interval of $\tau\sim11.9$~\textmu s, which is consequently adopted in all subsequent experiments. 

\subsection{Analytical model for logical relaxation rates}

The logical relaxation errors from $|1_\mathrm{L}\rangle$ to $|0_\mathrm{L}\rangle$ are quantified by initializing the logical qubit in $|1_\mathrm{L}\rangle$ and repetitively applying $M$ cycles of mid-circuit erasure detections at a fixed interval $\tau$. The postselected population in $|1_\mathrm{L}\rangle$ as a function of the total evolution time $t = MT$ reveals the logical relaxation rates within the codespace of the erasure qubit. Here, $T$ denotes the duration of a single cycle. The evolution of a single experimental cycle can be modeled as the composition of two quantum channels: free evolution ($\mathcal{E}_\mathrm{FE}$) followed by a mid-circuit erasure detection ($\mathcal{E}_\mathrm{ED}$).

The free evolution of an arbitrary cavity state $\rho$ (truncated to the Fock qutrit subspace spanned by $\{\ket{0}, \ket{1}, \ket{2}\}$) under the dominant single-photon loss at a rate of $\kappa_c$ is governed by the Lindblad master equation:
\begin{equation}
\frac{d\rho}{dt} = \frac{\kappa_c}{2}\left(2a\rho a^\dagger - a^\dagger a \rho - \rho a^\dagger a  \right).
\end{equation}
Over a finite interval $\tau$, this free time evolution can be described by a completely positive and trace-preserving map $\mathcal{E}_\mathrm{FE}(\rho) = \sum_{k=0}^{2}{E_k\rho E_k^\dagger}$, with Kraus operators $E_k=\sqrt{\frac{(1-e^{-\kappa_c\tau})^k}{k!}}e^{-\frac{\kappa_c\tau a^\dagger a}{2} }a^k$~\cite{michael2016}. Here, $k=0$ corresponds to the no-jump backaction and $k>0$ describes $k$-photon loss events.  

The mid-circuit erasure detection projects the cavity state onto the logical code space, effectively removing any population that leaked into $|1\rangle$. This process is modeled by the quantum channel: 
\begin{equation}
\mathcal{E}_{\mathrm{ED}}(\rho) = \frac{\mathcal{P}_\mathrm{C} \rho \mathcal{P}_\mathrm{C}^\dagger}{\operatorname{Tr}(\mathcal{P}_\mathrm{C} \rho \mathcal{P}_\mathrm{C}^\dagger)},
\end{equation}
where $\mathcal{P}_\mathrm{C}=|0_\mathrm{L}\rangle\langle 0_\mathrm{L}| + |1_\mathrm{L}\rangle\langle 1_\mathrm{L}|$ represents the projector onto the codespace. 

These two quantum channels together compose a full experimental cycle: 
\begin{equation}
\begin{aligned}
\mathcal{E}_{\mathrm{cycle}}(\rho) &= \mathcal{E}_\mathrm{ED} \circ \mathcal{E}_\mathrm{FE}(\rho)= \mathcal{E}_\mathrm{ED}(\mathcal{E}_\mathrm{FE}(\rho))\\
&\propto |0\rangle\langle 0|\rho|0\rangle\langle 0| + e^{-2\kappa_c\tau}|2\rangle\langle 2|\rho|2\rangle\langle 2| + (1-e^{-\kappa_c\tau})|0\rangle\langle 1|\rho|1\rangle\langle 0| + (1-e^{-\kappa_c\tau})^2|0\rangle\langle 2|\rho|2\rangle\langle 0|,
\end{aligned}
\end{equation}
where we have ignored the normalization factor. Starting from an initial state $\rho_0 = |2\rangle\langle 2|$, we can directly obtain the resulting cavity states after 1, 2, ... $M$ cycles as
\begin{equation}
\begin{aligned}
\mathcal{E}_{\mathrm{cycle}}(\rho_0) &\propto e^{-2\kappa_c\tau}|2\rangle\langle 2| + (1-e^{-\kappa_c\tau})^2|0\rangle\langle 0|,\\
\mathcal{E}_{\mathrm{cycle}}^{\circ 2}(\rho_0) &\propto e^{-4\kappa_c\tau}|2\rangle\langle 2| + (1-e^{-\kappa_c\tau})^2(1+e^{-2\kappa_c\tau})|0\rangle\langle 0|,\\
& ...,\\
\mathcal{E}_{\mathrm{cycle}}^{\circ M}(\rho_0) &\propto e^{-2M\kappa_c\tau}|2\rangle\langle 2| + (1-e^{-\kappa_c\tau})^2(1+e^{-2\kappa_c\tau}+...+e^{-2M\kappa_c\tau})|0\rangle\langle 0|
\end{aligned}
\end{equation}
Therefore, the corresponding population remaining in $|1_\mathrm{L}\rangle$ can be extracted analytically as 
\begin{equation}
\label{eq:P1L}
\begin{aligned}
P_{|1_\mathrm{L}\rangle}&=\langle 2| \mathcal{E}_\mathrm{cycle}^{\circ N}(\rho_0) |2\rangle = \frac{e^{-2M\kappa_c\tau}}{e^{-2M\kappa_c\tau} + (1-e^{-\kappa_c\tau})^2(1+e^{-2\kappa_c\tau}+...+e^{-2M\kappa_c\tau})} \\
&= \frac{e^{\kappa_c\tau}+1}{(e^{\kappa_c\tau}-1)e^{2M\kappa_c\tau}+2},
\end{aligned}
\end{equation}
with $M=t/T$.

In the long-time limit, consecutive single-photon loss or multi-photon loss events occurring within a single interval $\tau$ cannot be identified via the conventional parity measurement of the erasure check and manifest as undetectable logical errors. These detrimental effects would make the population in $|1_\mathrm{L}\rangle$ decays exponentially as $P_{|1_\mathrm{L}\rangle}\propto e^{-2\kappa_c\tau t/T}$. 

At short times, the dominant single-photon loss during the interval $\tau$ can be effectively detected via the erasure check, thereby significantly minimizing the decay of the logical population. By expanding the analytical expression in Eq.~\ref{eq:P1L} in a short-time limit, we could obtain a linear decay of $P_{|1_\mathrm{L}\rangle} \approx 1 - \gamma_\mathrm{int}t$ with an intrinsic logical relaxation rate $\gamma_\mathrm{int} = (\kappa_c\tau)^2/T$. 

The analytical expression described in Eq.~\ref{eq:P1L} accounts only for the dominant single-photon loss of the erasure qubit. However, several experimental imperfections exist during the idling and detection measurements, including ancilla errors, cavity thermal excitations, and detection-induced errors during the sequence. To account for these imperfections, we add an exponential decay term to Eq.~\ref{eq:P1L} to capture the residual relaxation rate within the codespace. The measured population in Fig.~3b in the main text is fitted to an effective model $P_{\ket{1_\mathrm{L}},\mathrm{eff}}(t) = A\, P_{\ket{1_\mathrm{L}}}(t)\, e^{-\gamma_{\mathrm{res}} t} + B$, where the fitting parameters $A$ and $B$ account for experimental imperfections and $\gamma_\mathrm{res}$ is the residual relaxation rate. Expanding this fitting function at short times gives a linear decay $\sim 1-(\gamma_\mathrm{int} + \gamma_\mathrm{res})t$, yielding the total effective logical relaxation rate $\gamma_\mathrm{tot} = \gamma_\mathrm{int} + \gamma_\mathrm{res}$.

\begin{figure}[b]
\includegraphics{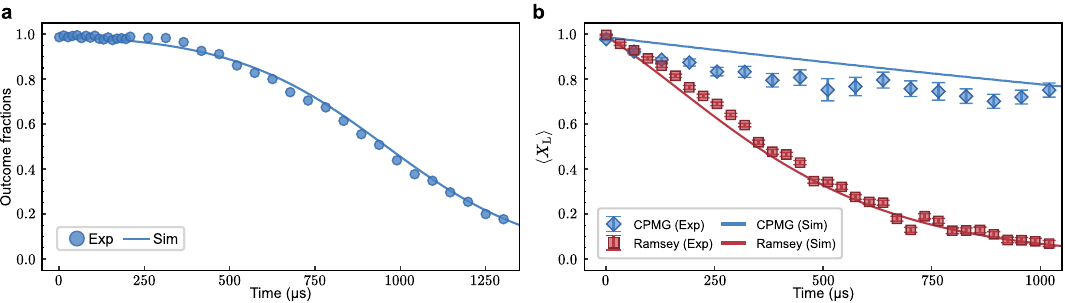}
\caption{Comparison between experiment data (symbols) and numerical simulation results (solid lines) for the logical relaxation experiment (\textbf{a}), and the logical Ramsey and CPMG experiments (\textbf{b}). The experimental data are same to that in Fig.~3 in the main text.
}
\label{fig:cpmg_vs_ramsey}
\end{figure}

\subsection{Numeric simulation}  
To further investigate the logical relaxation and dephasing experiments, we also perform time-domain numerical simulations using Qutip master equation solvers~\cite{qutip5}. In these simulations, we employ a comprehensive system Hamiltonian that includes cross-Kerr and higher-order self-Kerr interactions of the cavity-ancilla system. To account for pulse imperfections, we use pulse-level simulations for the numerically optimized pulses in initialization and gates, as well as for the pulses in the parity measurement process. The simulation of the ancilla readout also incorporates a phenomenological map that accounts for state assignment errors due to IQ-plane overlap~\cite{chou2024}. All experimentally measured parameters and decoherence times (Table~\ref{tab:parameter_sum}) are used in these simulations. Finally, the resulting system density matrix from the simulations is used to extract the logical state probabilities, allowing for a direct comparison with the experimental results.

Figure~\ref{fig:cpmg_vs_ramsey}a presents the simulation result for the logical relaxation experiment, which shows good agreement with experimental data, indicating the effectiveness of the error modeling in the simulation. We also present the simulation results for the logical Ramsey and CPMG-type experiments in Fig.~\ref{fig:cpmg_vs_ramsey}b. For the logical Ramsey experiment, the simulation also matches the measurement data. For the simulation of the logical CPMG-type experiment, we omit the pure dephasing time of the cavity to account for the elimination of low-frequency dephasing noises from the refocusing pulses. However, this simplification may slightly overestimate the remaining dephasing noises, leading to a minor discrepancy. Nevertheless, the overall trends in these simulations remain consistent with the corresponding experimental results.

\section{Logical gate error characterization}

In the main text, we characterize the logical gate errors of the erasure qubit using a randomized benchmarking (RB) experiment with interleaved mid-circuit erasure detections. This supplementary section provides further details on the RB experiment and presents additional process tomography results to clarify the logical error mechanisms during gate operations.

\subsection{Randomized benchmarking}

We first employ the Clifford-based RB method~\cite{PhysRevA.77.012307, PhysRevLett.106.180504} to quantify the performance of single-qubit logical gates of the erasure qubit. The RB sequence consists of the following steps: (i) apply $M$ random single-qubit logical Clifford gates to the initial state $|0_\mathrm{L}\rangle$; (ii) insert a mid-circuit erasure detection after each Clifford gate; (iii) append a recovery Clifford gate and a final erasure check to invert the whole sequence; (iv) measure the logical qubit population remaining in $|0_\mathrm{L}\rangle$ via photon-number-resolved measurement.

The single-qubit Clifford gates are constructed using virtual $Z$ rotations (realized by frame transformations~\cite{mckay2017}) and a physically implemented logical $X_{\mathrm{L}}/2$ gate with numerically optimized pulses of a duration $1.2$~\textmu s. The decomposition of the 24 single-qubit Clifford gates is summarized in Table~\ref{tab:gate_decomposition}.   

The postselection probability (i.e., the fraction of experimental shots with no erasure flagged) of all erasure detections after the Clifford gates decays exponentially with the number of Clifford gates $M$, revealing an erasure error probability of $9.77\times 10^{-2}$ per Clifford. The measured state probability remaining in $|0_\mathrm{L}\rangle$ after postselection is fitted to an exponential function $F(M) = Ap^M + B$, yielding a residual error of $r_\mathrm{Clif}=(1-p)/2=6.21\times 10^{-3}$ per Clifford. Given that each Clifford contains on average 2.17 $X_{\mathrm{L}}/2$ gates based on the previous decomposition (see Table.~\ref{tab:gate_decomposition}), we obtain average erasure error of $4.50\times 10^{-2}$ and residual gate error of $2.86\times 10^{-3}$ for the elementary logical $X_{\mathrm{L}}/2$ gate after postselection.

\begin{table}[b]
    \centering
    \caption{Decomposition of single-qubit Clifford group for the erasure qubit.}
    \renewcommand{\arraystretch}{1.5}
    \setlength{\tabcolsep}{10pt}
    \makebox[\textwidth][c]{
    \begin{tabular}{|l|l|l|}
        \hline
        \multicolumn{1}{|c|}{\textbf{Angle}} & 
        \multicolumn{1}{c|}{\textbf{Axis}} & 
        \multicolumn{1}{c|}{\textbf{Decomposition}} \\
        \hline
        --- & --- & ($I_\mathrm{L}$) \\
        $\pi$ & (1,0,0) & ($X_\mathrm{L}/2$) - ($X_\mathrm{L}/2$) \\
        $\pi$ & (0,1,0) & ($-Z_\mathrm{L}/2$) - ($X_\mathrm{L}/2$) - ($X_\mathrm{L}/2$) - ($Z_\mathrm{L}/2$) \\
        $\pi$ & (0,0,1) & ($-Z_\mathrm{L}/2$) - ($X_\mathrm{L}/2$) - ($X_\mathrm{L}/2$) - ($Z_\mathrm{L}/2$) - ($X_\mathrm{L}/2$) - ($X_\mathrm{L}/2$) \\
        $\pi/2$ & (1,0,0) & ($X_\mathrm{L}/2$) \\
        $\pi/2$ & (-1,0,0) & ($Z_\mathrm{L}$) - ($X_\mathrm{L}/2$) - ($Z_\mathrm{L}$) \\
        $\pi/2$ & (0,1,0) & ($-Z_\mathrm{L}/2$) - ($X_\mathrm{L}/2$) - ($Z_\mathrm{L}/2$) \\
        $\pi/2$ & (0,-1,0) & ($Z_\mathrm{L}/2$) - ($X_\mathrm{L}/2$) - ($-Z_\mathrm{L}/2$) \\
        $\pi/2$ & (0,0,1) & ($Z_\mathrm{L}$) - ($X_\mathrm{L}/2$) - ($Z_\mathrm{L}$) - ($-Z_\mathrm{L}/2$) - ($X_\mathrm{L}/2$) - ($Z_\mathrm{L}/2$) - ($X_\mathrm{L}/2$) \\
        $\pi/2$ & (0,0,-1) & ($Z_\mathrm{L}$) - ($X_\mathrm{L}/2$) - ($Z_\mathrm{L}$) - ($Z_\mathrm{L}/2$) - ($X_\mathrm{L}/2$) - ($-Z_\mathrm{L}/2$) - ($X_\mathrm{L}/2$) \\
        $\pi$ & (1,0,1) & ($X_\mathrm{L}/2$) - ($X_\mathrm{L}/2$) - ($Z_\mathrm{L}/2$) - ($X_\mathrm{L}/2$) - ($-Z_\mathrm{L}/2$) \\
        $\pi$ & (-1,0,1) & ($X_\mathrm{L}/2$) - ($X_\mathrm{L}/2$) - ($-Z_\mathrm{L}/2$) - ($X_\mathrm{L}/2$) - ($Z_\mathrm{L}/2$) \\
        $\pi$ & (0,1,1) & ($-Z_\mathrm{L}/2$) - ($X_\mathrm{L}/2$) - ($X_\mathrm{L}/2$) - ($Z_\mathrm{L}/2$) - ($X_\mathrm{L}/2$) \\
        $\pi$ & (0,-1,1) & ($-Z_\mathrm{L}/2$) - ($X_\mathrm{L}/2$) - ($X_\mathrm{L}/2$) - ($Z_\mathrm{L}/2$) - ($Z_\mathrm{L}$) - ($X_\mathrm{L}/2$) - ($Z_\mathrm{L}$) \\
        $\pi$ & (1,1,0) & ($X_\mathrm{L}/2$) - ($-Z_\mathrm{L}/2$) - ($X_\mathrm{L}/2$) - ($Z_\mathrm{L}/2$) - ($X_\mathrm{L}/2$) \\
        $\pi$ & (1,-1,0) & ($Z_\mathrm{L}$) - ($X_\mathrm{L}/2$) - ($Z_\mathrm{L}$) - ($-Z_\mathrm{L}/2$) - ($X_\mathrm{L}/2$) - ($Z_\mathrm{L}/2$) - ($Z_\mathrm{L}$) - ($X_\mathrm{L}/2$) - ($Z_\mathrm{L}$) \\
        $2\pi/3$ & (1,1,1) & ($-Z_\mathrm{L}/2$) - ($X_\mathrm{L}/2$) - ($Z_\mathrm{L}/2$) - ($X_\mathrm{L}/2$) \\
        $2\pi/3$ & (-1,1,1) & ($-Z_\mathrm{L}/2$) - ($X_\mathrm{L}/2$) - ($Z_\mathrm{L}/2$) - ($Z_\mathrm{L}$) - ($X_\mathrm{L}/2$) - ($Z_\mathrm{L}$)\\
        $2\pi/3$ & (1,-1,1) & ($Z_\mathrm{L}/2$) - ($X_\mathrm{L}/2$) - ($-Z_\mathrm{L}/2$) - ($X_\mathrm{L}/2$) \\
        $2\pi/3$ & (-1,-1,1) & ($Z_\mathrm{L}/2$) - ($X_\mathrm{L}/2$) - ($-Z_\mathrm{L}/2$) - ($Z_\mathrm{L}$) - ($X_\mathrm{L}/2$) -($Z_\mathrm{L}$) \\
        $-2\pi/3$ & (1,1,1) & ($Z_\mathrm{L}$) - ($X_\mathrm{L}/2$) - ($Z_\mathrm{L}$) - ($Z_\mathrm{L}/2$) - ($X_\mathrm{L}/2$) - ($-Z_\mathrm{L}/2$) \\
        $-2\pi/3$ & (-1,1,1) & ($X_\mathrm{L}/2$) - ($Z_\mathrm{L}/2$) - ($X_\mathrm{L}/2$) - ($-Z_\mathrm{L}/2$) \\
        $-2\pi/3$ & (1,-1,1) & ($Z_\mathrm{L}$) - ($X_\mathrm{L}/2$) - ($Z_\mathrm{L}$) - ($-Z_\mathrm{L}/2$) - ($X_\mathrm{L}/2$) - ($Z_\mathrm{L}/2$) \\
        $-2\pi/3$ & (-1,-1,1) & ($X_\mathrm{L}/2$) - ($-Z_\mathrm{L}/2$) - ($X_\mathrm{L}/2$) - ($Z_\mathrm{L}/2$) \\
        \hline
    \end{tabular}}
    \label{tab:gate_decomposition}
\end{table}

\subsection{Process tomography}

\begin{figure}[b]
    \begin{center}
        \includegraphics{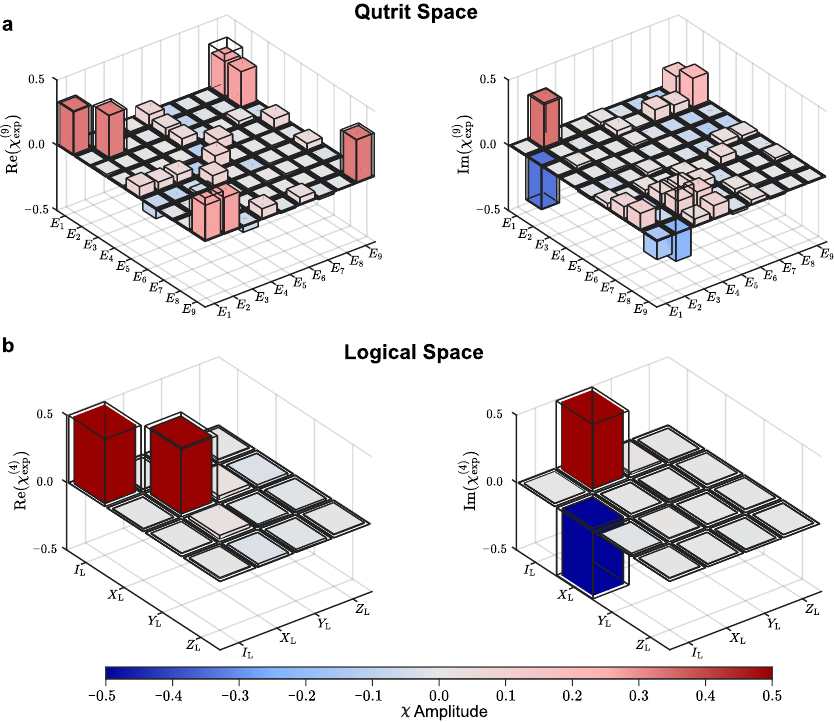}
    \end{center}
    \caption{Experimental process matrix of the logical $X_{\mathrm{L}}/2$ gate. \textbf{a} Bar charts of the real (left) and imaginary (right) parts of the full $9 \times 9$ process matrix for the achieved logical $X_{\mathrm{L}}/2$ gate in the Fock qutrit space. 
\textbf{b} Bar charts of the real (left) and imaginary (right) parts of the reduced $4 \times 4$ process matrix truncated into the logical subspace $\{\ket{0},\ket{2}\}$. 
       }
    \label{fig:process_tomo}
\end{figure}

To further assess the performance of the logical $X_{\mathrm{L}}/2$ gate on the erasure qubit, we perform quantum process tomography experiments within the three-level Fock qutrit subspace spanned by \{$|0\rangle$, $|1\rangle$, $|2\rangle$\}. In this experiment, we first initialize one of the nine states $\rho_\mathrm{ini} \in \{ \ket{0_{\mathrm{L}}},\ket{1_{\mathrm{L}}},\ket{+x_{\mathrm{L}}},\ket{-y_{\mathrm{L}}},\ket{1},\frac{\ket{0}+\ket{1}}{\sqrt{2}},\frac{\ket{0}-i\ket{1}}{\sqrt{2}},\frac{\ket{1}+\ket{2}}{\sqrt{2}}, \frac{\ket{1}-i\ket{2}}{\sqrt{2}} \}$, then apply the logical $X_{\mathrm{L}}/2$ gate, and finally measure the output cavity state via Wigner tomography to reconstruct the cavity density matrix $\rho_\mathrm{out}$ via MLE. With these nine initial states $\rho_\mathrm{ini}$ and the nine corresponding final states $\rho_\mathrm{out}$, we can obtain the process matrix $\chi_\mathrm{exp}^{(9)}$ of the logical $X_{\mathrm{L}}/2$ gate through the relation $\rho_\mathrm{out} = \sum_{m,n}{\chi_{mn}E_m\rho_\mathrm{ini}E_n^\dagger}$~\cite{Nielsen2010}, where the complete set of nine orthogonal basis operators within the qutrit subspace is chosen as
\begin{align}
\label{E_matrix}
	E_1 & = \sqrt{\frac{3}{2}}\begin{pmatrix} 1 & 0 & 0 \\ 0 & 0 & 0 \\ 0 & 0 & 1 \end{pmatrix},
	    & E_2 & = \sqrt{\frac{3}{2}}\begin{pmatrix} 0 & 0 & 1 \\ 0 & 0 & 0 \\ 1 & 0 & 0 \end{pmatrix},
	    & E_3 & = \sqrt{\frac{3}{2}}\begin{pmatrix} 0 & 0 & -1 \\ 0 & 0 & 0 \\ 1 & 0 & 0 \end{pmatrix}, \nonumber \\[6pt]
	E_4 & = \sqrt{\frac{3}{2}}\begin{pmatrix} 1 & 0 & 0 \\ 0 & 0 & 0 \\ 0 & 0 & -1 \end{pmatrix},
	    & E_5 & = \sqrt{\frac{3}{2}}\begin{pmatrix} 0 & 1 & 0 \\ 1 & 0 & 0 \\ 0 & 0 & 0 \end{pmatrix},
	    & E_6 & = \sqrt{\frac{3}{2}}\begin{pmatrix} 0 & -1 & 0 \\ 1 & 0 & 0 \\ 0 & 0 & 0 \end{pmatrix}, \\[6pt]
	E_7 & = \sqrt{\frac{3}{2}}\begin{pmatrix} 0 & 0 & 0 \\ 0 & 0 & 1 \\ 0 & 1 & 0 \end{pmatrix},
	    & E_8 & = \sqrt{\frac{3}{2}}\begin{pmatrix} 0 & 0 & 0 \\ 0 & 0 & -1 \\ 0 & 1 & 0 \end{pmatrix},
	    & E_9 & = \sqrt{3}\begin{pmatrix} 0 & 0 & 0 \\ 0 & 1 & 0 \\ 0 & 0 & 0 \end{pmatrix}. \nonumber
\end{align}

Figure~\ref{fig:process_tomo}a presents the reconstructed full process matrix $\chi_\mathrm{exp}^{(9)}$ of the logical $X_{\mathrm{L}}/2$ gate on the three-level Fock qutrit subspace. To evaluate the gate performance, we compute the normalized process fidelity $F_\chi = \left|\mathrm{Tr}\left(\chi_\mathrm{exp}\chi_\mathrm{ideal}^\dagger\right)\right|/\sqrt{\mathrm{Tr}\left(\chi_\mathrm{exp}\chi_\mathrm{exp}^\dagger\right) \mathrm{Tr}\left(\chi_\mathrm{ideal}\chi_\mathrm{ideal}^\dagger\right) }$~\cite{xu2018} to compare the experimental $\chi_\mathrm{exp}^{(9)}$ with the ideal process matrix $\chi_\mathrm{ideal}$ for the logical $X_\mathrm{L}/2$ gate. The relatively low value of the resulting $F_\chi = 0.51$ indicates a large amount of population transition errors associated with the leakage state $|1\rangle$ during the gate operation. 

To isolate the performance of the $X_\mathrm{L}/2$ gate only within the logical subspace, we extract the reduced $4\times 4$ process matrix $\chi_\mathrm{exp}^{(4)}$ within the codespace by involving the process with only $|0\rangle$ and $|2\rangle$ states and discarding any operations acting on the leakage state $|1\rangle$. The extracted reduced process matrix within the logical subspace is shown in Fig.~\ref{fig:process_tomo}b, yielding a normalized process fidelity of 0.998, which is consistent with the residual gate error extracted from the RB experiment.  

The discrepancy between the full-space and logical subspace process fidelities originates primarily from pulse imperfections that induce unwanted residual evolution of the erasure state $|1\rangle$ during the gate operation. Nevertheless, such control-induced unwanted evolutions associated with the erasure state can be effectively eliminated and converted into detectable erasures by performing postselection against erasure errors from the mid-circuit erasure detection, preventing these control-induced erasure errors from propagating into the logical subspace.

\normalem
%